%% file: bare_jrnl.tex
\pdfoutput=1

\documentclass[journal]{IEEEtran}
\usepackage{booktabs}   
\usepackage{longtable}  
\usepackage{amsmath}
\usepackage{placeins}  
\usepackage{graphicx}
\graphicspath{{figures/}} 
\usepackage{placeins}
\usepackage{stfloats}   
\usepackage{array}      

\newcounter{taskctr}
\newcolumntype{N}{>{\stepcounter{taskctr}\arabic{taskctr}}r} 

\ifCLASSINFOpdf
\else
\fi
\begin{document}
%
\title{COMPASS: A Multi-Dimensional Benchmark for Evaluating Code Generation in Large Language Models}
%
%
%

\author{James~Meaden\\
        Michał~Jarosz\\
        Piotr~Jodłowski\\
        Grigori~Melnik
\thanks{All authors are affiliated with Codility.\\
Corresponding author: James Meaden; james.meaden@codility.com}%
}

\maketitle

\begin{abstract}
Current code generation benchmarks focus primarily on functional correctness while overlooking two critical aspects of real-world programming: algorithmic efficiency and code quality. We introduce COMPASS (COdility's Multi-dimensional Programming ASSessment), a comprehensive evaluation framework that assesses code generation across three dimensions: correctness, efficiency, and quality. COMPASS consists of 50 competitive programming problems from real Codility competitions, providing authentic human baselines from 393,150 submissions. Unlike existing benchmarks that treat algorithmically inefficient solutions identically to optimal ones provided they pass test cases, COMPASS systematically evaluates runtime efficiency and code quality using industry-standard analysis tools. Our evaluation of three leading reasoning-enhanced models, Anthropic Claude Opus 4, Google Gemini 2.5 Pro, and OpenAI O4-Mini-High, reveals that models achieving high correctness scores do not necessarily produce efficient algorithms or maintainable code. These findings highlight the importance of evaluating more than just correctness to truly understand the real-world capabilities of code generation models. COMPASS serves as a guiding framework, charting a path for future research toward AI systems that are robust, reliable, and ready for production use.\end{abstract}

\begin{IEEEkeywords}
code generation, large language models, benchmarking, software engineering assessment, AI model evaluation, COMPASS
\end{IEEEkeywords}

%
\IEEEpeerreviewmaketitle

\section{Introduction}
%
%
%
%
\IEEEPARstart{L}{arge} language models (LLMs) have achieved remarkable proficiency in code
generation, yet current evaluation benchmarks predominantly assess only functional correctness
through test case execution \cite{chen2021codellm,austin2021programsynthesis,jimenez2023swebench}.
This narrow focus fails to capture essential aspects of real-world software development where code
must not only work correctly but also perform efficiently and maintain high quality standards.
Existing benchmarks such as HumanEval \cite{chen2021codellm}, MBPP \cite{austin2021programsynthesis},
and HackerRank-ASTRA \cite{xing2025astra} evaluate models based on functional correctness,
measuring whether generated code produces expected outputs for given test cases. While these
benchmarks have been valuable for measuring code generation capabilities, they overlook critical
aspects of professional programming: the ability to write code that is maintainable and runs efficiently. In real-world software development, a solution that passes functional tests but
runs in exponential time or violates coding best practices would be considered inadequate, yet
current benchmarks would score it as perfect.

This evaluation gap is problematic given the increasing deployment of LLMs in production environments where code efficiency and quality directly impact user experience, system security and scalability, and long-term maintainability \cite{letouzey2012sqale,chidamber1994metrics}. Recent reasoning-enhanced models demonstrate sophisticated problem-solving capabilities, yet the field lacks a comprehensive understanding of whether this enhanced reasoning translates to better algorithmic efficiency and code quality.

To address these limitations, we introduce COMPASS (COdility's Multi-dimensional Programming ASSessment), a benchmark that evaluates code generation in three critical dimensions: correctness, efficiency, and quality. COMPASS consists of 50 competitive programming problems from real coding competitions hosted by Codility, providing comprehensive percentile rankings of LLMs against human performance baselines from 393{,}150 submissions. We developed COMPASS to serve as a navigational aid for the field, enabling researchers and professionals to pinpoint where current models truly stand in terms of real-world performance, and to chart a clear path forward toward robust, efficient, and maintainable AI-driven code generation.

\section{Related Work and Motivation}\label{sec:related}

\subsection{Limitations of Current Benchmarks}
Most existing benchmarks reduce evaluation to a single axis: functional correctness.
HumanEval \cite{chen2021codellm} helped popularize this approach, and later benchmarks such as APPS \cite{hendrycks2021apps}, SWE\mbox{-}bench \cite{jimenez2023swebench}, DS\mbox{-}1000 \cite{lai2022ds1000}, and HackerRank\mbox{-}ASTRA \cite{xing2025astra} largely preserve this narrow focus, even when applied to more complex, real-world tasks.

This leads to two critical blind spots:
\begin{itemize}
\item \textbf{Efficiency:} A brute-force $\mathcal{O}(n^{3})$ implementation is treated
as equivalent to an optimal $\mathcal{O}(n \log n)$ algorithm as long as both pass test cases designed to assess only code functionality. This fails to reflect real-world software development, where efficiency determines feasibility and scalability.
  \item \textbf{Quality:} Key aspects such as maintainability, readability, modularity, and adherence to best practices—critical for long-term productivity—are largely ignored in prominent evaluations. Current benchmarks prioritize syntactic correctness while overlooking principles of sustainable and scalable software engineering \cite{letouzey2012sqale,chidamber1994metrics}.
\end{itemize}

\subsection{The Need for Multi-Dimensional Assessment}
The importance of efficiency and quality is far from theoretical. Competitive programming platforms such as \emph{Codeforces} and \emph{TopCoder} impose strict runtime and memory constraints, while decades of software engineering research emphasize multi-dimensional metrics for assessing code quality.
However, current code generation benchmarks fall short of capturing these essential dimensions. By relying primarily on pass@k metrics, they reduce software development to a binary pass/fail task---overlooking the efficiency, quality, and sustainability that define real engineering value. The result is a skewed portrayal of model capabilities, risking overestimation of their readiness for real-world use. This reveals a critical disconnect between what benchmarks measure and what modern software development truly demands.

\section{COMPASS Benchmark Design}\label{sec:design}

COMPASS includes 50 coding challenges selected from Codility’s public contests held between 2011 and 2021. These problems span a wide range of algorithmic domains and difficulty levels, and are accompanied by large-scale human performance data, enabling more nuanced and multidimensional model evaluation. Participants could complete these challenges using any of 25$+$ mainstream programming languages, including C, C++, Python, Java, JavaScript, Go, Rust, Swift, Kotlin, PHP, Ruby, C\#, TypeScript, and others. This diversity ensures that solutions reflect a broad range of programmer preferences and constraints. In this initial study, we restricted model-generated solutions to Python~3 to control for variance in performance outcomes across language implementations. Future research using the COMPASS benchmark will expand this analysis to include a broader set of programming languages.

Each problem in COMPASS has been attempted by thousands of participants in real programming competitions, resulting in 393{,}150 human submissions. This large-scale dataset provides empirical insight into common patterns of failure and variation in problem-solving approaches under time-constrained conditions. Problems were selected to ensure diversity in topic coverage and to expose variation in reasoning depth, algorithmic efficiency, and implementation robustness—dimensions not captured by correctness-only benchmarks.

To contextualize model performance, we report aggregate statistics from the full set of historical human submissions (see Appendix~\ref{app:human-norms}) along with percentile scores for each model against human-based norms. Overall benchmark data from human submissions are:
\begin{itemize}
  \item \textbf{Average human mean score:} 23.31\% (range: 11.00\%--40.40\%).
  \item \textbf{Average human median score:} 9.35\% (range: 0.00\%--44.00\%).
  \item \textbf{Average human standard deviation:} 31.08.
  \item \textbf{Average human skew:} $+1.36$.
\end{itemize}

The positive average skew in human scores indicates a long tail of low-scoring attempts, evidence of high difficulty and common pitfalls such as inefficient solutions, misinterpreted constraints, or failure to handle edge cases. In short: these are hard problems, even for experienced programmers.

By anchoring the benchmark in large-scale human data from open competitions, COMPASS offers a high-signal, multi-dimensional evaluation of code generation systems, assessing not just whether a model produces a working solution, but \emph{how} it performs under the same pressure-tested conditions that reveal meaningful variation in human skill. Appendix~\ref{app:human-norms} shows the human performance baselines on all 50 COMPASS problems.

\subsection{Multi-Dimensional Evaluation Framework}\label{sec:framework}
COMPASS evaluates each model solution across three independent dimensions: \emph{functional correctness}, \emph{computational efficiency}, and \emph{code quality}. This structure reflects how software is evaluated in the real world: an examination of not just \emph{whether} it works, but \emph{how well} it works and \emph{how maintainable} it is. Below we describe each pillar of our multidimensional evaluation framework.

\subsubsection{Correctness}
Each problem includes a comprehensive test suite spanning:
\begin{itemize}
  \item Basic functionality
  \item Edge cases
  \item Corner cases
\end{itemize}
Correctness is scored as the percentage of test cases passed, providing a more granular view of solution robustness compared to frameworks that rely on binary pass/fail metrics, i.e.,
\begin{equation}
C_{\text{corr}} = 100 \cdot \frac{N_{\text{passed}}}{N_{\text{total}}}.
\label{eq:correctness}
\end{equation}
where $N_{\text{passed}}$ is the number of tests passed and $N_{\text{total}}$ is the total number of tests.

\subsubsection{Efficiency}
Efficiency test cases are explicitly designed to stress algorithmic scalability. Each efficiency test case includes large input sizes near the upper bounds of the problem’s constraints. These cases are evaluated under strict runtime thresholds derived from efficient, expert-crafted reference solutions with known optimal time and space complexity. The efficiency score is calculated as the percentage of efficiency test cases passed. This dimension separates solutions that are merely correct from those that are correct \emph{and} scalable.

\subsubsection{Quality}
Beyond correctness and speed, COMPASS evaluates how well-written the code is, using static analysis via CodeScene, an industrial-grade code quality platform. Well-written code is easy to understand, modify, and extend, thus minimizing technical debt and supporting long-term maintainability \cite{letouzey2012sqale,chidamber1994metrics}.

Using Codility’s internal CodeScene configuration, COMPASS analyzes each solution across multiple dimensions of code quality, including: cyclomatic and cognitive complexity, function length, and nesting depth (complexity); code duplication, cohesion, method design, parameter structure, and naming conventions (maintainability); and adherence to language-specific idioms, detection of anti-patterns, and overall structural quality (best practices). Each solution receives a composite code quality score (1–10), normalized to a 1–100 scale for consistency across evaluation metrics. In addition to overall code quality scores, we investigate the frequency of specific code quality issues per code generation model. 

\section{Experimental Setup}\label{sec:setup}

\subsection{Model Selection}
We evaluate three state-of-the-art reasoning-enhanced models representing the current frontier in code generation: Anthropic Claude Opus 4, Google Gemini 2.5 Pro, OpenAI O4-Mini-High. 

\subsection{Evaluation Protocol}

\paragraph{Sampling Strategy.}
We employ $k=64$ independent samples per model--problem combination, using the model’s default temperature setting ($T=1.0$). This sampling strategy balances computational cost with statistical rigor, enabling stable estimates of central tendency and variance. A sample size of 64 allows for reliable mean comparisons across models and problems, while capturing meaningful variability in the nuances of model code generation.

\paragraph{Prompt Variation.}
To investigate how prompt framing might influence model behavior, we tested four distinct prompt types, each designed to emphasize a different optimization objective during code generation. Prompts were appended to the start of each coding challenge. 

\medskip
\noindent\textbf{Prompt 1: Neutral (Control).}
This baseline prompt provided no additional guidance beyond solving the problem “effectively”:
\begin{quote}\small\itshape
Below is a programming challenge. Your task is to write code that solves the problem. You will use only the Python programming language, version 3. You may not install additional libraries.

\medskip
Write a solution that effectively solves the problem as described.
\end{quote}

\noindent\textbf{Prompt 2: Correctness-Optimized.}
This version emphasized the importance of passing all test cases:
\begin{quote}\small\itshape
Below is a programming challenge. Your task is to write code that solves the problem. You will use only the Python programming language, version 3. You may not install additional libraries.

\medskip
Prioritize writing code that passes all test cases, including edge cases and uncommon input patterns.
\end{quote}

\noindent\textbf{Prompt 3: Efficiency-Optimized.}
This prompt instructed the model to focus on computational efficiency:
\begin{quote}\small\itshape
Below is a programming challenge. Your task is to write code that solves the problem. You will use only the Python programming language, version 3. You may not install additional libraries.

\medskip
Prioritize writing code that runs efficiently, especially on large or complex inputs.
\end{quote}

\noindent\textbf{Prompt 4: Dual Objective.}
This prompt combined the priorities of both correctness and efficiency:
\begin{quote}\small\itshape
Below is a programming challenge. Your task is to write code that solves the problem. You will use only the Python programming language, version 3. You may not install additional libraries.

\medskip
Prioritize writing code that passes all test cases while also running efficiently, especially on large or complex inputs.
\end{quote}

This four-way prompt design enabled us to systematically explore potential trade-offs between correctness, efficiency, and code quality as a function of prompt conditioning.

\section{Results and Analysis}\label{sec:results}

\subsection{Multidimensional Assessment Validation}
Before analyzing model performance, we first validated our multidimensional evaluation framework to ensure that correctness, efficiency, and quality provide complementary and non-redundant signals. This step is essential for establishing the value of a multidimensional scoring approach, in line with best practices in human performance assessment \cite{campbell1990performance}. To do so, we conducted correlation analyses and principal components analysis (PCA). The goal was to confirm that each dimension contributes unique information to the overall performance composite (i.e., the aggregated score resulting from correctness, efficiency, and quality), rather than simply reflecting the same underlying domain of performance.

\subsubsection{Correlation Analyses}
We computed pairwise correlations across the full dataset of model submissions ($N=3{,}200$ per correlation). Correctness and efficiency were moderately correlated ($r=.655$, $p{<}.001$), suggesting that more correct completions tend to also execute more efficiently. In contrast, code quality was not correlated with correctness ($r=.089$, $p{<}.001$) or efficiency ($r=.022$, $p{<}.001$). These low effect sizes indicate that the code-quality metric captures a distinct dimension of performance, orthogonal to writing code that merely works or runs quickly.

These findings support the structure of our framework: all three dimensions reflect meaningfully different aspects of software performance, and should be retained as independent axes in model evaluation.

\begin{table}[!htbp]
\caption{Correlation Between Performance Dimensions}
\label{tab:dim-corr}
\centering
\begin{tabular}{lccc}
\toprule
& \textbf{Correctness} & \textbf{Efficiency} & \textbf{Quality} \\
\midrule
\textbf{Correctness} & 1     & 0.655 & 0.089 \\
\textbf{Efficiency}  & 0.655 & 1     & 0.022 \\
\textbf{Quality}     & 0.089 & 0.022 & 1     \\
\bottomrule
\end{tabular}
\end{table}

\FloatBarrier  

To better understand the relationship between dimensions in this multi-dimensional performance framework, we examined correlation matrices separately by model. While the general pattern of moderate correctness–efficiency alignment and code–quality independence holds overall, notable variations emerge across models.

\noindent\textbf{O4-Mini-High (Table~\ref{tab:o4})}. O4-Mini-High shows a strong correlation between correctness and efficiency ($r=.809$), suggesting that correct outputs are frequently accompanied by efficient execution, and vice versa. However, correlations with code quality are negligible ($r=-.014$ with correctness; $r=.005$ with efficiency), indicating that code quality is independent of the model’s ability to write correct and efficient code.

\begin{table}[!htbp]
\caption{O4-Mini-High Correlations}
\label{tab:o4}
\centering
\begin{tabular}{lccc}
\toprule
& \textbf{Correctness} & \textbf{Efficiency} & \textbf{Quality} \\
\midrule
\textbf{Correctness} & 1     & 0.809 & -0.014 \\
\textbf{Efficiency}  & 0.809 & 1     & 0.005  \\
\textbf{Quality}     & -0.014 & 0.005 & 1     \\
\bottomrule
\end{tabular}
\end{table}
\FloatBarrier

\noindent\textbf{Gemini 2.5 Pro (Table~\ref{tab:Gemini 2.5 Pro})}. Gemini 2.5 Pro shows a moderate-to-strong correctness–efficiency correlation ($r=.655$) and a negligible correctness–quality correlation ($r=.067$). While there is a low efficiency–quality correlation ($r=.224$), this relationship is notably higher than with O4-Mini-High, suggesting a slightly stronger tendency for Gemini 2.5 Pro to write higher-quality code when it writes highly efficient code.

\begin{table}[!htbp]
\caption{Gemini 2.5 Pro Correlations}
\label{tab:Gemini 2.5 Pro}
\centering
\begin{tabular}{lccc}
\toprule
& \textbf{Correctness} & \textbf{Efficiency} & \textbf{Quality} \\
\midrule
\textbf{Correctness} & 1     & 0.655 & 0.067 \\
\textbf{Efficiency}  & 0.655 & 1     & 0.224 \\
\textbf{Quality}     & 0.067 & 0.224 & 1     \\
\bottomrule
\end{tabular}
\end{table}
\FloatBarrier

\noindent\textbf{Claude Opus 4 (Table~\ref{tab:Claude Opus 4})}. Claude Opus 4 presents a lower correctness–efficiency correlation ($r=.503$) relative to the other models. Notably, it exhibits the strongest relationship between correctness and code quality ($r=.359$), with a weak but non-zero efficiency–quality correlation ($r=.165$). While these effect sizes are still relatively small, they suggest that Claude Opus 4 may generate more structurally sound code when it produces correct solutions.

\begin{table}[!htbp]
\caption{Claude Opus 4 Correlations}
\label{tab:Claude Opus 4}
\centering
\begin{tabular}{lccc}
\toprule
& \textbf{Correctness} & \textbf{Efficiency} & \textbf{Quality} \\
\midrule
\textbf{Correctness} & 1     & 0.503 & 0.359 \\
\textbf{Efficiency}  & 0.503 & 1     & 0.165 \\
\textbf{Quality}     & 0.359 & 0.165 & 1     \\
\bottomrule
\end{tabular}
\end{table}
\FloatBarrier

\noindent\textbf{In summary,}
\begin{itemize}
  \item When O4-Mini-High writes functional (i.e., correct) code, that code is also very likely to be efficient, but may or may not be high quality.
  \item When Gemini 2.5 Pro writes functional (i.e., correct) code, that code is likely to be efficient, but may or may not be high quality
  \item When Claude Opus 4 writes functional (i.e., correct) code, that code is only somewhat likely to be efficient, but is also more likely to be high quality.
\end{itemize}

Across all three models, none of the evaluation dimensions—correctness, efficiency, or code quality—are so highly correlated as to indicate redundancy. While correctness and efficiency tend to be moderately-highly correlated, particularly for O4-Mini-High and Gemini 2.5 Pro, code quality remains independent from both. This pattern holds not only in the aggregate analysis but also within each model individually. These results provide initial empirical support for treating correctness, efficiency, and quality as complementary, non-overlapping dimensions of performance in code generation evaluation.

\subsubsection{Principal Components Analysis}
To support the conceptual and empirical separation of correctness, efficiency, and code quality, we conducted a PCA. While our evaluation framework treats these as distinct dimensions by design, PCA allows us to test whether they in fact capture independent sources of variation in model behavior, rather than overlapping aspects of a single latent factor.

The first principal component (PC1) explained 55.4\% of the total variance and loaded strongly on both correctness (0.705) and efficiency (0.699), suggesting a shared performance dimension related to basic task completion. In contrast, code quality showed a minimal loading on PC1 (0.118), indicating it contributes little to this common factor. The second component (PC2) accounted for an additional 33.1\% of the variance and loaded almost exclusively on code quality (0.990), with negligible contributions from correctness and efficiency. This confirms that code quality represents a statistically distinct and orthogonal dimension, not captured by correctness or efficiency alone. The third component (PC3) explained the remaining 11.4\% of variance and showed opposing loadings for correctness (0.708) and efficiency ($-$0.702), suggesting a residual tradeoff between these metrics. This may reflect cases where models produce correct but inefficient solutions (e.g., brute-force approaches) or, conversely, generate efficient code that fails to handle all test cases. While PC3 accounts for a smaller portion of variance, it highlights nuanced patterns that would not be evident from correlation analysis alone.

\begin{figure}[!htbp]
  \centering
  \includegraphics[width=\linewidth]{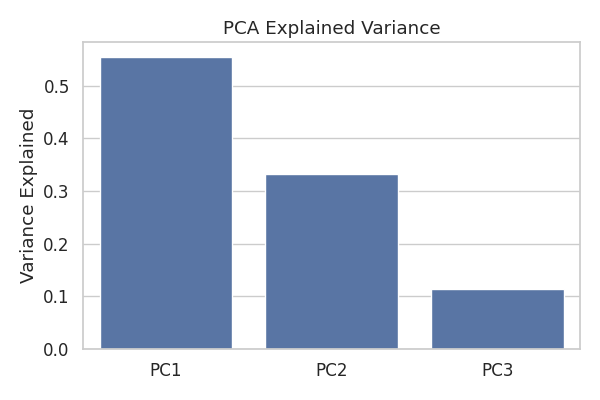} 
  \caption{PCA explained variance across components.}
  \label{fig:pca}
\end{figure}

\begin{table}[!htbp]
  \caption{PCA Loadings}
  \label{tab:pca-loadings}
  \centering
  \begin{tabular}{lrrr}
    \toprule
    & \textbf{PC1} & \textbf{PC2} & \textbf{PC3} \\
    \midrule
    \textbf{Correctness} & 0.705 & -0.033 & 0.708 \\
    \textbf{Efficiency}  & 0.699 & -0.134 & -0.702 \\
    \textbf{Quality}     & 0.118 &  0.990 & -0.071 \\
    \bottomrule
  \end{tabular}
\end{table}
\FloatBarrier

Taken together, these results validate the structure of our multi-dimensional framework. Correctness and efficiency load strongly on the same principal component, indicating they reflect a shared performance dimension. However, their opposing loadings on a third component suggest that each still captures distinct problem-solving tendencies—such as favoring brute-force correctness over efficiency, or vice versa. Meanwhile, code quality loads almost exclusively on a separate component, highlighting its role as an independent signal not captured by correctness or efficiency. This dimensional structure confirms that all three metrics contribute complementary information, reinforcing the need for a benchmark like COMPASS that evaluates model performance across multiple, independently meaningful axes.

\subsection{Overall Performance Summary}\label{sec:overall}

With the validity of the multidimensional scoring framework established, we now summarize model performance across the three COMPASS evaluation dimensions—correctness, efficiency, and quality—along with the resulting composite scores (i.e., the average of the three dimensions).

Across all 50 COMPASS problems, the three models achieved high average correctness scores, indicating strong baseline capabilities in producing syntactically valid and functionally correct code. However, differences across models become more apparent when considering efficiency and code quality, as well as variability across runs. Table VI shows overall model performance scores (the numbers in the table, and all similar tables that follow, represent median / mean and standard deviation values).

\begin{table*}[!b]
\caption{Overall model performance scores (Median / Mean $\pm$ SD for each model on the four COMPASS metrics).}
\label{tab:overall-model-scores}
\centering
\small
\begin{tabular}{lcccc}
\toprule
\textbf{Model} & \textbf{Correctness} & \textbf{Efficiency} & \textbf{Quality} & \textbf{Composite} \\
\midrule
\texttt{Claude-Opus-4} & 100 / 72.2 $\pm$ 36.5 & 22.2 / 35.4 $\pm$ 39.1 & 93.8 / 92.3 $\pm$ 6.2 & 66.7 / 66.1 $\pm$ 23.8 \\
\texttt{Gemini-2.5-Pro}         & 100 / 93.8 $\pm$ 20.3 & 100 / 85.4 $\pm$ 30.3  & 94.5 / 93.2 $\pm$ 5.9 & 97.9 / 90.4 $\pm$ 17.1 \\
\texttt{O4-Mini-High}           & 100 / 95.6 $\pm$ 17.4 & 100 / 93.0 $\pm$ 21.5  & 93.7 / 89.2 $\pm$ 12.9 & 97.5 / 92.3 $\pm$ 15.5 \\
\bottomrule
\end{tabular}
\end{table*}

\FloatBarrier

Claude Opus 4 exhibited the lowest overall performance and the highest variability. While its code quality score was strong (mean of 92.3), its efficiency score was substantially lower (mean of 35.4), and its correctness scores were lower and less consistent (mean of 72.2, SD of 36.5). These limitations are reflected in its low mean composite score of 66.1 and standard deviation (SD) of 23.8.

O4-Mini-High delivered the most consistently high performance in both correctness (mean of 95.6) and efficiency (mean of 93) dimensions. Though scored the lowest in quality (mean of 89.2). Its mean composite score of 92.3 was the highest across the three models evaluated. O4-Mini-High also showed low variability in its scores, having the smallest SD for all dimensions except quality.

Gemini 2.5 Pro also performed strongly across all three dimensions, with correctness (mean of 93.8), efficiency (mean of 85.4), and the highest code quality (mean of 93.2) along with the lowest variability in code quality (SD of 5.9). Its overall composite mean score was 90.4.

\subsection{Dimension-Specific Analysis}\label{sec:dim-specific}

\subsubsection{Correctness Analysis}
\begin{figure}[!htbp]
  \centering
  \includegraphics[width=\linewidth]{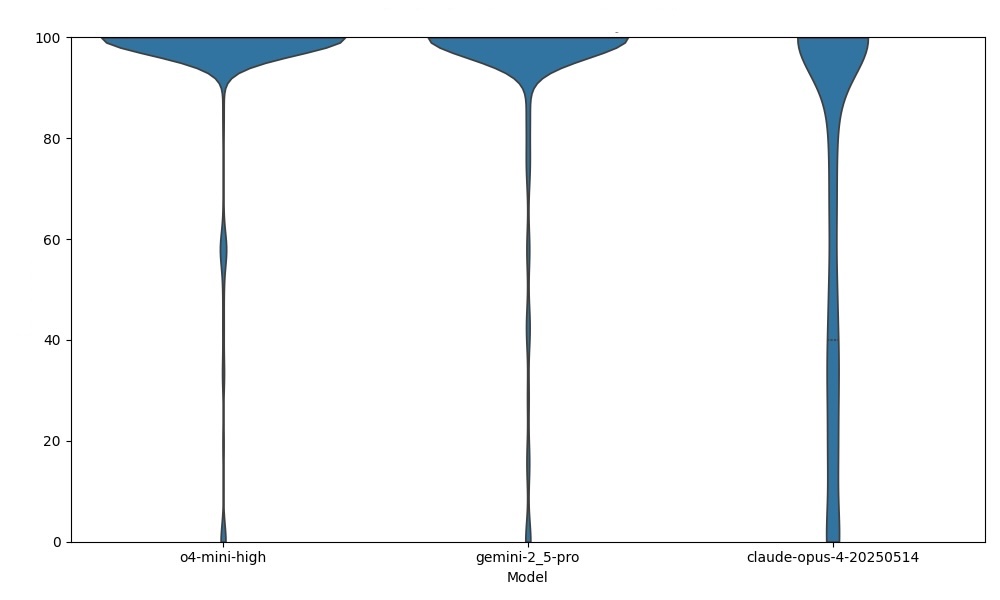}
  \caption{Distribution of correctness scores by model across 50 tasks.}
  \label{fig:correctness-violin}
\end{figure}

The violin plot in Fig.~\ref{fig:correctness-violin} shows the distribution of correctness scores for each
model across all 50 benchmark tasks. While all three models demonstrate a large concentration of scores
near the upper end of the scale, differences in distributional shape and spread are notable. The O4-Mini-High model displays a highly peaked distribution near 100\%, with minimal variance
and few low-performing outliers, indicating that this model solves nearly every task correctly and
consistently. Gemini 2.5 Pro also exhibits a strong mode near 100\%, but with a slightly
broader spread, suggesting occasional drops in performance on specific tasks. In contrast,
Claude Opus 4 shows a markedly wider and flatter distribution, with a substantial number of
scores clustering well below 80\% and a long tail extending to zero. This reveals much more variable
performance, with several tasks where Claude Opus 4 did not provide correct and functional code. Overall,
the plot highlights O4-Mini-High's consistency in writing highly functional code, Gemini 2.5 Pro's
similar capabilities with slightly less consistency, and Claude Opus 4's capability to write correct and
functional code but with very low consistency (Appendix~B contains the per-task efficiency score statistics for each model).

\FloatBarrier

\subsubsection{Efficiency Analysis}
The distribution of efficiency scores, reflecting runtime efficiency and solution speed, reveals important contrasts across the three models. Both O4-Mini-High and Gemini 2.5 Pro, once again, show tightly clustered distributions near the upper limit of the scale, indicating a very high consistency in producing highly efficient code. Most outputs from these models executed quickly and without timeouts, reflected in their high means (93.0 for O4-Mini-High; 85.4 for Gemini 2.5 Pro).

\begin{figure}[!htbp]
  \centering
  \includegraphics[width=\linewidth]{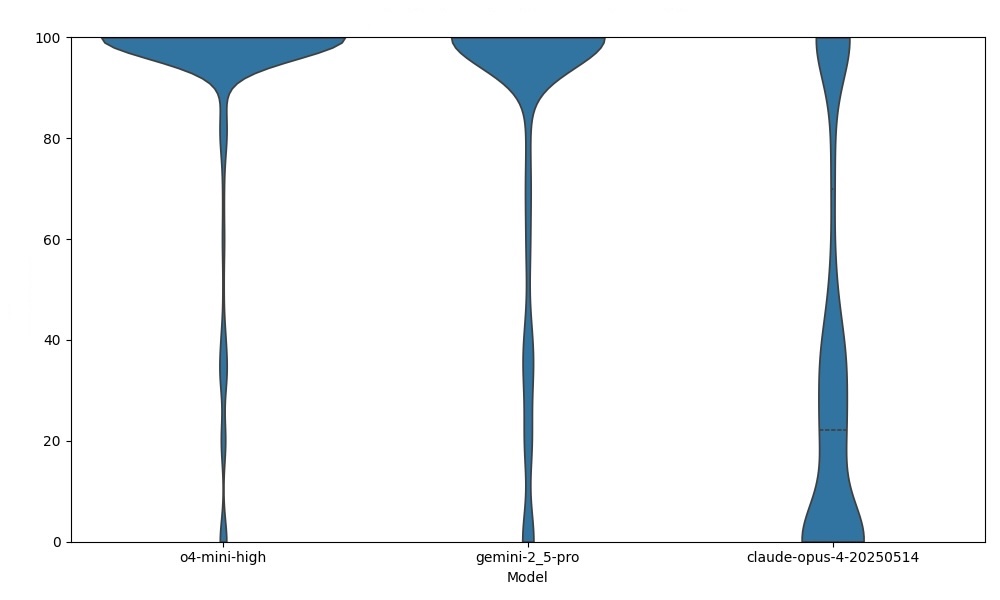}
  \caption{Distribution of efficiency scores by model across 50 tasks.}
  \label{fig:efficiency-violin}
\end{figure}

In contrast, Claude Opus 4 exhibited a wide and skewed distribution of efficiency scores. Although it occasionally produced efficient solutions, many solutions timed out or were otherwise inefficient, resulting in a mean efficiency score of 35.4. This is visible in the violin plot (Fig.~\ref{fig:efficiency-violin}), where Claude Opus 4’s distribution shows a high density near the lower end of the efficiency scale, reflecting repeated failures to generate efficient code. These findings suggest that while all models tend to produce functional (i.e., correct) code (see previous section), there are substantial differences in how efficient that code is. This distinction has important implications for deployment in real-world environments (Appendix~C contains the per-task efficiency score statistics for each model).

\FloatBarrier

\subsubsection{Code Quality Analysis}
\begin{figure}[!htbp]
  \centering
  \includegraphics[width=\linewidth]{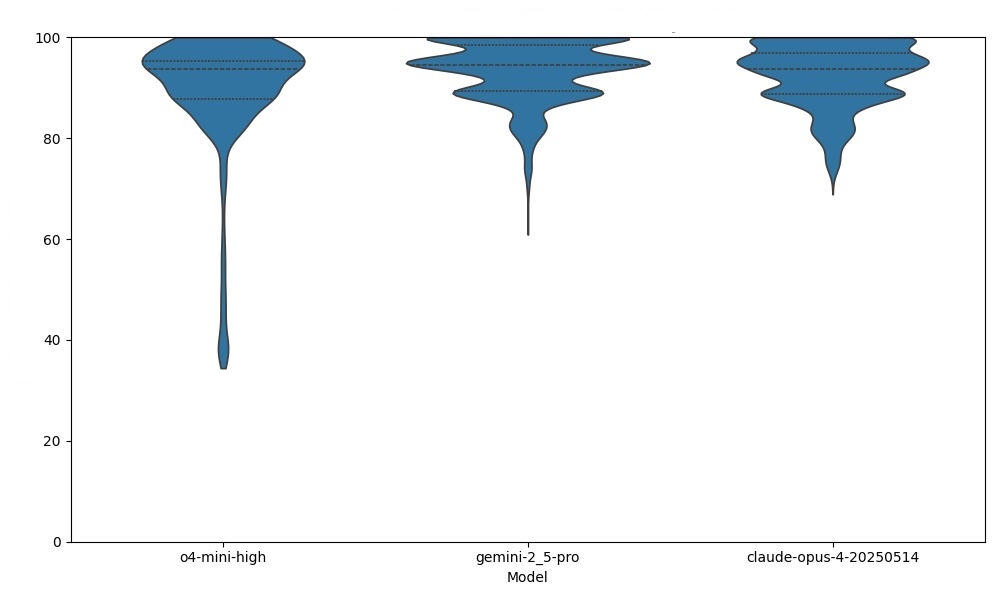}
  \caption{Distribution of code quality scores by model across 50 tasks.}
  \label{fig:quality-violin}
\end{figure}

While correctness and efficiency evaluate whether a model produces a working and efficient solution, code quality reflects how readable, maintainable, and idiomatic (i.e., stylistically aligned with best practices in the language) that solution is. High-quality code is easier to debug, extend, and deploy, making it a critical dimension in practical software development. To assess this, we used CodeScene’s evaluation method based on automated linters and code structure heuristics. These scores capture adherence to style conventions, use of clear variable names, modularity, and other best practices commonly valued in software engineering. Figure~\ref{fig:quality-violin} displays the distribution of code quality scores across all models (Appendix~D contains the per-task quality score statistics for each model).

The violin plot reveals that all three models demonstrate relatively similar and consistently high performance on code quality. The distributions are tightly clustered between 80\% and 100\%, with long, dense central bodies and narrow tails. This suggests that, regardless of differences in correctness or efficiency, all three models are capable of producing clean, idiomatic Python code with minimal syntactic or stylistic flaws.

O4-Mini-High shows a skew toward the higher end of the distribution, with a smaller amount of perfect scores and fewer outputs below 80\%. Gemini 2.5 Pro and Claude Opus 4 show consistently high scores mostly above 80\%. This convergence in code quality suggests that modern LLMs, even when struggling to produce correct or efficient solutions, have internalized robust stylistic norms and best practices. It also confirms the statistical finding from our PCA: code quality represents a largely independent axis of model behavior. The practical implication is that even when models fail to produce working or fast code, the outputs may still be readable and salvageable, making them potentially useful for collaborative workflows where human review and modification follow automated generation.

\FloatBarrier

In addition to reviewing model code quality scores, examining the number and type of code quality issues provides a complementary perspective on model performance. Table VII shows the number of specific problems flagged across model submissions, offering more granular insight into the typical nature and frequency of code quality concerns for each model. While all three models had a similar number of code quality issues, O4-Mini-High showed the highest mean issue count per submission (2.048), followed by Claude Opus 4 (1.950) and Gemini 2.5 Pro (1.874). The median number of issues per submission was consistent across models at 2.

\begin{table}[!htbp]
\caption{Summary of Code Quality Issues By Model}
\label{tab:quality-issues-summary}
\centering
\small
\begin{tabular}{lccc}
\toprule
\textbf{Model} & \textbf{Total} &
\textbf{\shortstack{Mean Issue\\Per Submission}} &
\textbf{\shortstack{Median Issue\\Per Submission}} \\
\midrule
\texttt{Claude-Opus-4} & 6237 & 1.950 & 2 \\
\texttt{Gemini-2.5-Pro}         & 5998 & 1.874 & 2 \\
\texttt{O4-Mini-High}           & 6555 & 2.048 & 2 \\
\bottomrule
\end{tabular}
\end{table}
\FloatBarrier

Following the high-level issue count analysis, we examined the specific types of code quality issues flagged across model outputs (Table VIII). While the total issue counts differ only modestly between models, the types of issues reveal more about each model’s stylistic and structural tendencies. This breakdown highlights whether problems are primarily cosmetic, structural, or symptomatic of deeper design flaws—insights that are critical for developers evaluating downstream maintainability.

\noindent\textbf{Issue profile by category.}
As shown in Table~\ref{tab:quality-issues-specific}, all three models exhibit a high proportion of issues in three categories: \emph{Bumpy Road Ahead}, \emph{Deep, Nested Complexity}, and \emph{Complex Method}---all structural markers of cognitive load and refactoring difficulty. O4-Mini-High had the highest rate of ``Bumpy Road'' issues (74.5\%), indicating a tendency to embed multiple logic segments in single functions. Claude Opus 4 (67.1\%) and Gemini 2.5 Pro (65.6\%) also showed a high rate of this issue.

\begin{table*}[!t]
\caption{Specific Code Quality Issues by Model}
\label{tab:quality-issues-specific}
\centering
\scriptsize
\setlength{\tabcolsep}{5pt}
\renewcommand{\arraystretch}{1.1}
\begin{tabular}{
  >{\raggedright\arraybackslash}p{0.19\textwidth}  
  >{\raggedright\arraybackslash}p{0.24\textwidth}  
  >{\raggedleft\arraybackslash}p{0.15\textwidth}  
  >{\raggedleft\arraybackslash}p{0.15\textwidth}  
  >{\raggedleft\arraybackslash}p{0.15\textwidth}} 
\toprule
\textbf{Code Quality Issue} & \textbf{Description} &
\textbf{Claude-Opus-4} & \textbf{Gemini-2.5-Pro} & \textbf{O4-Mini-High} \\
\midrule
Bumpy Road Ahead & Functions with multiple chunks of nested logic, making code harder to read, refactor, and reason about. & 2154 (67.1\%) & 2100 (65.6\%) & 2353 (74.5\%) \\
Deep, Nested Complexity & Excessive control structure nesting increases cognitive load and is strongly linked to bugs. & 1383 (43.2\%) & 848 (26.5\%) & 1127 (35.2\%) \\
Complex Method & High cyclomatic complexity means the function has too many logical branches, reducing readability and maintainability. & 2088 (65.2\%) & 2250 (69.7\%) & 2380 (74.4\%) \\
Large Method & Overly long functions make the code harder to read and understand. & 174 (5.4\%) & 399 (12.5\%) & 222 (6.9\%) \\
Complex Conditional & Branches with multiple logical conditions reduce clarity and should be simplified or encapsulated. & 375 (11.7\%) & 220 (6.9\%) & 344 (10.8\%) \\
Global Conditions & Complex logic outside of functions should be refactored into named functions to improve structure. & 3 (0.1\%) & 64 (2.0\%) & 35 (1.1\%) \\
Excess Number of Function Arguments & Too many function parameters suggest low cohesion or missing abstractions. & 68 (2.1\%) & 124 (3.9\%) & 64 (2.0\%) \\
Overall Code Complexity & Measures average logical branching per function; higher values suggest more testing and complexity. & 0 (0.0\%) & 4 (0.1\%) & 0 (0.0\%) \\
Code Duplication & Identical or near-identical code blocks across functions increase maintenance overhead and reduce clarity. & 0 (0.0\%) & 9 (0.3\%) & 0 (0.0\%) \\
\bottomrule
\end{tabular}
\end{table*}
\FloatBarrier

\noindent\textbf{Deep nesting.}
Claude Opus 4 showed the highest incidence of deep nesting (43.2\%), suggesting frequent use of complex, multi-level control flow. This aligns with its greater variability in code quality scores. In contrast, O4-Mini-High (35.2\%) and Gemini 2.5 Pro (26.5\%) tended to generate simpler, flatter logic structures.

\noindent\textbf{Complex methods.}
O4-Mini-High had the highest rate of \emph{Complex Method} issues (74.4\%), followed by Gemini 2.5 Pro (69.7\%) and Claude Opus 4 (65.2\%). This suggests that, despite O4-Mini-High and Gemini 2.5 Pro’s strengths in modularity and clarity, they still tend to generate methods that are densely packed with logic and contain numerous branches.

\noindent\textbf{Other categories.}
Less frequent categories such as \emph{Large Method}, \emph{Complex Conditional}, and \emph{Excess Number of Function Arguments} still reveal model-level differences. Gemini 2.5 Pro had more than twice as many large methods (12.5\%) as Claude Opus 4 (5.4\%) or O4-Mini-High (6.9\%). Claude Opus 4 had more complex conditionals (11.7\%) than Gemini 2.5 Pro (6.9\%), reinforcing its tendency toward cognitively demanding logic.

\noindent\textbf{Rare issues.}
Rarely flagged issues such as \emph{Code Duplication}, \emph{Global Conditions}, and \emph{Overall Code Complexity} were minimal across all models, indicating general adherence to abstraction and reuse best practices.

\noindent\textbf{Takeaway.}
In sum, while all three models demonstrate broadly similar total issue volumes, their underlying patterns of complexity differ in meaningful ways. O4-Mini-High frequently embeds multiple logic segments and dense branching within single methods, despite its overall strengths in clarity. Claude Opus 4’s outputs are marked by deeper nesting and more complex conditionals, reflecting a style that may increase cognitive load during review and maintenance. Gemini 2.5 Pro stands out for producing larger methods, suggesting occasional lapses in decomposition despite otherwise strong modularity. These distinctions matter: they reveal that even when models pass functional tests, their structural tendencies can introduce hidden costs for long-term maintainability and engineering effort.

\subsubsection{Composite Score Analysis}
\begin{figure}[!htbp]
  \centering
  \includegraphics[width=\linewidth]{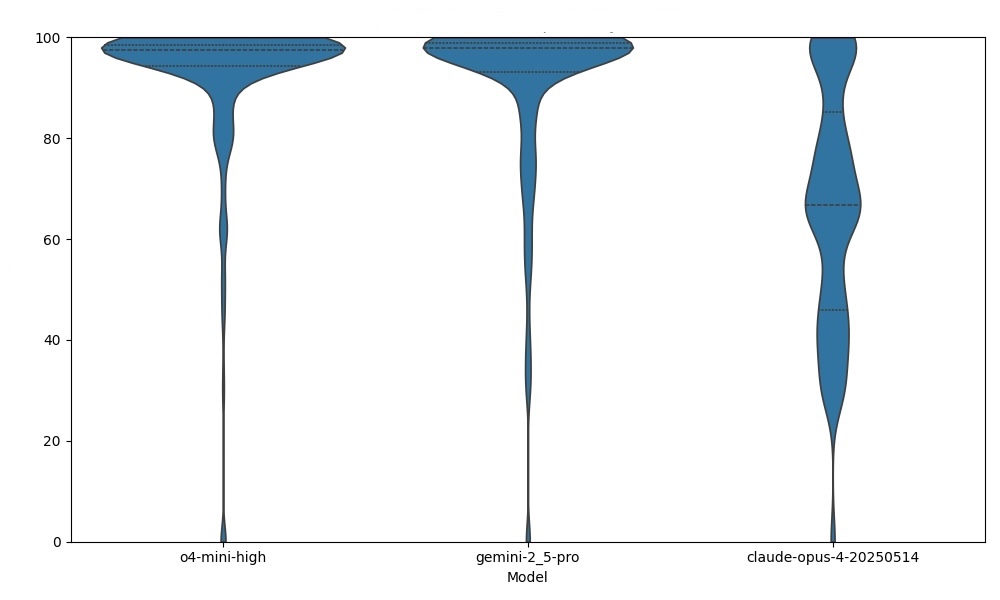}
  \caption{Distribution of composite scores by model across 50 tasks.}
  \label{fig:composite-violin}
\end{figure}

To provide a comprehensive view of model performance, we computed a composite score for each task by
equally weighting the three evaluation dimensions: correctness, efficiency, and quality. This score
reflects a model’s ability to generate not only correct, but also efficient and high-quality code.
Figure~\ref{fig:composite-violin} displays the distribution of these composite scores for each model
across all 50 benchmark tasks.

O4-Mini-High stands out for its consistent and top-tier performance, with the
vast majority of outputs achieving composite scores above 90\%. Its narrow, top-heavy distribution
indicates that it almost always delivers code that is correct, fast, and clean, with few outliers or
low-performing cases. The consistency across dimensions translates to highly predictable performance,
a major strength in production environments.

Gemini 2.5 Pro shows a similar pattern, with scores also tightly clustered at the upper end of the
scale. Its slightly wider spread and greater presence of mid-range scores suggest a small but noticeable
drop in consistency compared to O4-Mini-High, though overall performance remains strong.

Claude Opus 4, by contrast, exhibits a dramatically wider and flatter distribution. Its scores span a
broad range, with a significant proportion of tasks falling below the 80\% mark and a wider tail extending
toward zero. This reflects the model’s uneven performance across dimensions, particularly its frequent
struggles with efficiency.

\FloatBarrier

In sum, the COMPASS benchmark reveals that O4-Mini-High and Gemini 2.5 Pro are both highly performant
general-purpose code generators, each with slight advantages in either mean score or variability
depending on the task. Claude Opus 4, however, demonstrates inconsistent and often low composite performance,
largely due to low efficiency scores, potentially limiting its reliability in production contexts (Appendix~E contains the per-task composite score statistics for each model).

\subsection{Human Baseline Comparison}\label{sec:human-baselines}
To contextualize model performance against human programmers, we benchmarked each model’s combined
correctness and efficiency score relative to human participants in a series of Codility competitive
programming challenges. This analysis was limited to only correctness and efficiency scores as the
human benchmark data did not contain scores for code quality. These comparisons are based on
393{,}150 real-world human submissions.

For each task, we estimated the human-equivalent percentile rank of the model’s score by calculating
a $z$-score using the formula:
\begin{equation}
Z = \frac{X - \mu}{\sigma},
\label{eq:zscore}
\end{equation}
where $X$ is the model’s score, $\mu$ is the human mean, and $\sigma$ is the human standard deviation.
This $z$-score was then converted to a percentile using the cumulative distribution function (CDF) of
the standard normal distribution. For tasks with skewed or multimodal human performance data, these
estimates should be interpreted with caution (Appendix F contains the full
task-level data).

\begin{table}[!htbp]
\caption{Model Performance (Correctness and Efficiency) Against Human Benchmarks}
\label{tab:model-vs-human}
\centering
\small
\begin{tabular}{lcc}
\toprule
\textbf{Model} & \textbf{Mean Percentile} & \textbf{SD Percentile} \\
\midrule
\texttt{Claude-Opus-4}  & 76.580 & 19.478 \\
\texttt{Gemini2.5-pro} & 96.280 &  6.673 \\
\texttt{O4-Mini-High}   & 97.840 &  3.599 \\
\bottomrule
\end{tabular}
\end{table}
\FloatBarrier

Table~\ref{tab:model-vs-human} reports the mean human-equivalent percentile for each model, along
with the SD of those percentiles across tasks. O4-Mini-High achieved the highest average percentile rank
(97.840), followed closely by Gemini 2.5 Pro (96.280), while Claude Opus 4’s percentile rank was lowest (76.580).
The SD percentile reflects how much a model’s relative standing fluctuates across tasks: Claude Opus 4 shows
the greatest variation (SD $=$ 19.478), meaning its performance, compared to humans, varies
substantially depending on the task. In contrast, O4-Mini-High (SD $=$ 3.599) and Gemini 2.5 Pro (SD $=$ 6.673)
display far more consistent rankings, suggesting they outperform human participants at a more stable
level across tasks.

\subsubsection{Consistency Across Tasks}
While prior sections explored how model performance varies across tasks, here we examine within-task
consistency—how reliably a model produces similar results when solving the same problem multiple
times. For each task, we computed the SD of each model’s correctness, efficiency, and quality scores
across repeated generations. A lower mean SD indicates more consistent behavior. We also report the
SD of those SDs, capturing how much this consistency itself varies from task to task.

\begin{table}[!htbp]
\caption{Model Consistency: Per-Task Score Variability}
\label{tab:consistency}
\centering
\small
\begin{tabular}{l l cc}
\toprule
\textbf{Model} & \textbf{\shortstack{Performance\\Dimension}} & \textbf{Mean SD} & \textbf{SD of SDs} \\
\midrule
\texttt{O4-Mini-High}   & Correctness & 8.427  & 11.558 \\   
\texttt{O4-Mini-High}   & Efficiency  & 10.331 & 12.089 \\   
\texttt{O4-Mini-High}   & Quality     & 3.724  &  2.830 \\   

\texttt{Gemini-2.5-Pro} & Correctness & 10.560 & 12.082 \\
\texttt{Gemini-2.5-Pro} & Efficiency  & 13.312 & 14.003 \\
\texttt{Gemini-2.5-Pro} & Quality     &  3.153 &  1.728 \\

\texttt{Claude-Opus-4}  & Correctness & 17.366 & 14.165 \\
\texttt{Claude-Opus-4}  & Efficiency  & 15.895 & 13.753 \\
\texttt{Claude-Opus-4}  & Quality     &  3.435 &  1.797 \\
\bottomrule
\end{tabular}
\end{table}
\FloatBarrier

\noindent\textbf{Consistency summary.}
Table~\ref{tab:consistency} shows the average and variability of model performance consistency across repeated runs. O4-Mini-High exhibits the lowest mean standard deviations for correctness and efficiency, suggesting more stable behavior when attempting the same task multiple times. Claude Opus 4 shows notably higher variability—particularly in correctness—while Gemini 2.5 Pro generally falls in between.

\noindent\textbf{Magnitude of variability.}
While the direction of these differences aligns with earlier findings, the magnitude is not trivial. Even small standard deviations, which represent fluctuation in model performance, can impact user trust and complicate downstream evaluation or deployment decisions.

\noindent\textbf{Composite consistency score.}
To synthesize consistency across dimensions, we computed a composite SD for each task by aggregating the model’s correctness, efficiency, and code-quality SDs. Averaging these across tasks yields a single consistency score per model. O4-Mini-High had the lowest mean composite SD (7.49), indicating the most stable performance across repeated generations. Claude Opus 4 showed the highest variability (12.23), with Gemini 2.5 Pro again falling in between (9.01).

\begin{table}[!htbp]
\caption{Overall Model Consistency}
\label{tab:overall-consistency}
\centering
\small
\begin{tabular}{lcc}
\toprule
\textbf{Model} &
\textbf{\shortstack{Mean Composite\\SD}} &
\textbf{\shortstack{SD of Composite\\SDs}} \\
\midrule
\texttt{Claude-Opus-4} & 12.23 & 8.00 \\
\texttt{Gemini-2.5-Pro}         &  9.01 & 8.74 \\
\texttt{O4-Mini-High}           &  7.49 & 7.89 \\
\bottomrule
\end{tabular}
\end{table}
\FloatBarrier

We also report the SD of these composite SDs to assess whether consistency itself was stable across tasks. O4-Mini-High again showed the most uniform behavior (SD $=$ 7.89), while Gemini 2.5 Pro’s consistency varied the most from task to task (SD $=$ 8.74). Claude Opus 4 showed high variability in both metrics. Together, these results reinforce the finding that O4-Mini-High is not only strong on average performance but also more reliable and consistent, producing stable outputs across repeated runs and different tasks.

\subsection{Prompting Effects}\label{sec:prompting-effects}
To assess whether prompt phrasing influences model performance, we tested four distinct prompting
conditions across tasks: neutral, correctness\mbox{-}focused, efficiency\mbox{-}focused, and dual\mbox{-}focused
(emphasizing both correctness and efficiency). Each model received the same task under each of these
prompt framings, allowing us to isolate the effect of prompt intent on output quality.

\begin{table*}[!b]
\caption{Aggregated Prompting Effects}
\label{tab:prompting-effects}
\centering
\small
\begin{tabular}{lcccc}
\toprule
\textbf{Prompt type} & \textbf{Correctness} & \textbf{Efficiency} & \textbf{Quality} & \textbf{Composite} \\
\midrule
neutral     & 100 / 87.356 (28.042) & 100 / 70.966 (40.432) & 94 / 91.594 (9.411) & 96 / 82.894 (22.517) \\
correctness & 100 / 87.235 (27.891) & 100 / 70.293 (40.694) & 94 / 91.462 (9.548) & 96 / 82.647 (22.634) \\
efficiency  & 100 / 87.235 (28.293) & 100 / 72.357 (39.711) & 93.8 / 91.587 (8.692) & 96.2 / 83.343 (22.665) \\
dual        & 100 / 86.677 (28.691) & 100 / 71.509 (40.232) & 93.8 / 91.765 (8.654)  & 96.1 / 82.936 (29.901) \\
\bottomrule
\end{tabular}
\end{table*}
\FloatBarrier

Table~\ref{tab:prompting-effects} shows that prompt phrasing had only modest effects on model
performance at this aggregated level of analysis, but the observed differences were directionally
consistent with the intent of each prompt. For example, \emph{correctness} prompts produced slightly
higher average correctness scores (87.535) with the lowest variability (SD $=$ 27.891), and
\emph{efficiency} prompts led to the highest mean efficiency (72.357) with the lowest SD (39.711).
Similarly, \emph{dual}\mbox{-}focused prompts yielded the highest average code quality (91.765), while
\emph{neutral} prompts resulted in the most stable composite performance score (SD $=$ 22.517).

Although these differences are small in magnitude, they follow predictable patterns—indicating that
even lightweight prompt framing can subtly steer model behavior. It’s important to note that the
prompts used in this study were deliberately minimal: brief, high\mbox{-}level instructions rather than
detailed or example\mbox{-}driven prompts. It is plausible that more elaborate prompting strategies could
lead to stronger differentiation in model performance.

To better understand how individual models respond to prompt framing, we next examine the effects of
each prompt type on performance metrics for each model independently.

O4-Mini-High exhibited high and consistent performance across all prompt types, with minimal variation in response to prompt framing. Importantly, these small variations did not follow a consistent pattern aligned with prompt intent—for instance, the efficiency prompt did not clearly yield the highest efficiency, nor did the correctness prompt lead to the highest correctness. This suggests that O4-Mini-High is largely robust to prompt phrasing, and that its outputs are stable and strong regardless of minor differences in prompt framing (See Table XIII).

Gemini 2.5 Pro showed a generally stable pattern across all prompt types, but the results did not consistently reflect the intended emphasis of each prompt. The efficiency-focused prompt produced the highest efficiency score (86.171), aligning with its objective, but also yielded the highest scores for correctness (94.701). This suggests that while the efficiency prompt may have provided a clearer or more effective signal, the other prompt framings had minimal differentiating effect. Overall, Gemini 2.5 Pro appears largely robust to prompt variation, with high performance and low variability across prompt types (See Table XIV). 

In contrast to the other models, Claude Opus 4 was more responsive to prompt framing, with larger absolute differences across conditions (Table~\ref{tab:Claude Opus 4-prompting}). Correctness improved from 70.865 (efficiency prompt) to 73.155 (correctness prompt), and efficiency rose from 33.557 (correctness prompt) to 37.204 (efficiency prompt)---still below the other models, but consistently improved when given explicit guidance. Composite scores showed a similar pattern, increasing from 65.378 (neutral) to 67.403 (dual). Code quality, however, remained stable and high across prompts. These results suggest that Claude Opus 4 benefits more noticeably from prompt effects, and that targeted phrasing can meaningfully influence its output quality and consistency, though improvements still remain modest overall (See Table XV).

\section{Discussion}\label{sec:discussion}
The results presented in this paper challenge a central assumption embedded in current code generation benchmarks: that functional correctness alone is a sufficient proxy for real\mbox{-}world programming ability. By systematically evaluating correctness, efficiency, and code quality across a high\mbox{-}difficulty, human\mbox{-}anchored benchmark, COMPASS reveals that this assumption no longer holds---if it ever did.


\begin{table*}[!t]
\caption{O4-Mini-High Prompting Effects}
\label{tab:o4-prompting}
\centering
\small
\begin{tabular}{lcccc}
\toprule
\textbf{Prompt type} & \textbf{Correctness} & \textbf{Efficiency} & \textbf{Quality} & \textbf{Composite} \\
\midrule
neutral     & 100 / 95.964 (16.186) & 100 / 93.734 (20.256) & 93.5 / 88.771 (13.377) & 97.5 / 92.703 (13.574) \\
correctness & 100 / 95.604 (17.240) & 100 / 92.636 (21.788) & 93.7 / 88.801 (13.836) & 97.5 / 92.033 (15.333) \\
efficiency  & 100 / 96.140 (16.101) & 100 / 93.697 (20.414) & 93.8 / 89.890 (12.177) & 97.5 / 92.985 (14.470) \\
dual        & 100 / 94.692 (19.807) & 100 / 92.109 (23.407) & 93.5 / 89.396 (12.256) & 97.5 / 91.508 (18.054) \\
\bottomrule
\end{tabular}
\end{table*}

\vspace{-6pt}  

\begin{table*}[!t]
\caption{Gemini 2.5 Pro Prompting Effects}
\label{tab:Gemini 2.5 Pro-prompting}
\centering
\small
\begin{tabular}{lcccc}
\toprule
\textbf{Prompt type} & \textbf{Correctness} & \textbf{Efficiency} & \textbf{Quality} & \textbf{Composite} \\
\midrule
neutral     & 100 / 93.596 (19.831) & 100 / 85.411 (30.163) & 94.8 / 93.415 (5.862) & 97.9 / 90.601 (17.275) \\
correctness & 100 / 93.846 (20.039) & 100 / 84.685 (30.938) & 94.8 / 93.122 (5.916) & 97.9 / 90.255 (17.463) \\
efficiency  & 100 / 93.996 (19.019) & 100 / 86.171 (29.195) & 94.4 / 92.958 (6.072) & 97.9 / 90.937 (16.929) \\
dual        & 100 / 92.752 (22.343) & 100 / 85.179 (30.885) & 94.8 / 93.270 (5.856) & 97.8 / 89.898 (19.148) \\
\bottomrule
\end{tabular}
\end{table*}

\vspace{-6pt}  

\begin{table*}[!t]
\caption{Claude Opus 4 Prompting Effects}
\label{tab:Claude Opus 4-prompting}
\centering
\small
\begin{tabular}{lcccc}
\toprule
\textbf{Prompt type} & \textbf{Correctness} & \textbf{Efficiency} & \textbf{Quality} & \textbf{Composite} \\
\midrule
neutral     & 100 / 72.110 (36.806) & 20 / 33.753 (38.389) & 93.8 / 92.584 (6.379) & 66.4 / 65.378 (24.012) \\
correctness & 100 / 73.155 (36.397) & 20 / 33.557 (38.494) & 93.8 / 92.441 (6.106) & 66.7 / 65.652 (23.723) \\
efficiency  & 100 / 70.865 (37.135) & 25 / 37.204 (39.770) & 92.4 / 91.859 (6.265) & 66.3 / 66.107 (24.462) \\
dual        & 100 / 72.586 (35.778) & 25 / 37.238 (39.625) & 93.8 / 92.499 (6.047) & 66.7 / 67.403 (22.777) \\
\bottomrule
\end{tabular}
\end{table*}

\FloatBarrier  

While all three models evaluated (O4-Mini-High, Gemini 2.5 Pro, and Claude Opus 4) demonstrate impressive capabilities in generating correct code, only O4-Mini-High consistently delivered solutions that were also efficient and maintainable. Gemini 2.5 Pro followed closely, with similarly high correctness and slightly stronger code quality, but with occasional inefficiencies. Claude Opus 4, while capable of generating structurally sound code, displayed significantly higher variability and frequent inefficiencies, resulting in unpredictable and often suboptimal performance.

Crucially, these differences were not always apparent through correctness scores alone. Models that scored highly on functional correctness sometimes failed to produce efficient or well\mbox{-}structured solutions---outcomes that would raise red flags in real\mbox{-}world development. This disconnect underscores the practical limitations of benchmarks that treat software development as a binary pass/fail exercise. In production settings, an inefficient algorithm that passes tests can still cause unacceptable delays, costs, or scaling bottlenecks. Similarly, unmaintainable code---even if correct---can incur long\mbox{-}term costs in debugging, onboarding, and system reliability. The COMPASS multi\mbox{-}dimensional evaluation benchmark is designed to makes these hidden tradeoffs visible.

The value of a multi\mbox{-}dimensional evaluation framework is further reinforced by the orthogonality of its core metrics. Correlation and PCA analyses confirm that correctness, efficiency, and quality capture distinct axes of model behavior. While correctness and efficiency are moderately correlated (especially for O4-Mini-High) code quality remains largely independent. This means that a model’s ability to write readable, modular, idiomatic code cannot be inferred from whether it solves a problem correctly or quickly. These findings validate the need for multidimensional benchmarks: collapsing performance into a single metric obscures key subtleties that matter to software engineers and organizations.

The consistency analyses also reveal critical nuances. O4-Mini-High exhibited not only the highest average performance but also the most stable behavior across repeated runs and across tasks. Claude Opus 4, by contrast, fluctuated significantly, particularly in correctness and efficiency, raising concerns about reliability. Even small standard deviations can translate to substantial differences in outcome quality, especially when LLMs are used as coding assistants in workflows that expect dependable behavior. Composite consistency scores show that some models produce radically different outputs for the same task under the same conditions.

Prompt framing effects were subtle but informative. While small in magnitude, the directional alignment between prompt intent and performance outcomes suggests that model behavior can be steered, even with minimal prompting. Claude Opus 4, in particular, showed noticeably greater responsiveness to prompt framing---achieving higher correctness scores under correctness\mbox{-}oriented prompts and better efficiency under efficiency\mbox{-}oriented prompts. This pattern indicates that less performant models may benefit more from explicit guidance. In contrast, O4-Mini-High and Gemini 2.5 Pro displayed minimal variation across prompt types, suggesting a higher degree of robustness but also less room for gains from basic prompt tuning. These findings highlight an opportunity for more sophisticated methods of instruction, such as dynamic prompting, contextual tuning, or reinforcement learning from efficiency and quality objectives, not just correctness.

\noindent Collectively, these findings mark a turning point in how we evaluate AI code generation systems. The field can no longer afford to assess model performance through the narrow lens of functional correctness alone. Real software development is multi-dimensional. The benchmarks should be too.

\subsection*{Implications for Practice and Research}
For researchers, COMPASS provides a tool to evaluate not just \emph{whether} models can solve problems, but \emph{how} they solve them and at what cost. For practitioners, it offers insight into which models are production-ready, which require close supervision, and where the tradeoffs lie. For model developers, it highlights new optimization targets: computational efficiency, structural soundness, and performance stability.

Just as importantly, COMPASS sets a new standard for scientific rigor in benchmarking. By grounding evaluations in competitive programming tasks from real contests—supported by large-scale human baselines and robust statistical validation—it enables comparisons that are both empirically defensible and practically meaningful. While these tasks are not direct replicas of industry scenarios, they expose key model limitations in efficiency, robustness, and code quality under realistic constraints.

\subsection*{Future Directions}
This initial release of COMPASS focuses on Python solutions to algorithmic problems with well-defined performance constraints. Future work will expand the benchmark in several directions:
\begin{itemize}
  \item \textbf{Language expansion:} Add support for additional programming languages to enable broader and more representative model comparisons.
  \item \textbf{Task diversity:} Introduce more tasks that reflect realistic, real-life programming scenarios—including multi-file, project-based challenges that better approximate modern software development workflows.
  \item \textbf{Model coverage:} Expand evaluations to include a wider range of models, including open-source alternatives, to ensure broad benchmarking relevance.
  \item \textbf{Prompt complexity:} Incorporate a wider range of prompt types, varying in structure, specificity, and sophistication, to better understand how prompt design influences model behavior.
\end{itemize}

Most urgently, we call on the broader research and engineering communities to stop treating syntactic correctness as a proxy for software engineering capability. With models now reaching and exceeding human-level performance in certain correctness tasks, the question is no longer “Can they write code that works?”—but “Can they write code that lasts?”

COMPASS is not just a benchmark—it is a call to realign our standards with the realities of software development. The future of AI-assisted programming depends on it.

\appendices

\onecolumn
\section{Human Performance Baselines}\label{app:human-norms}

\setlength{\tabcolsep}{6pt}
\scriptsize
The table below summarizes human performance across the 50 COMPASS benchmark tasks, based on 393{,}150 real-world submissions from Codility competitive programming contests. For each task, we report the mean, median, standard deviation (SD), skewness, and total number of submissions.
\renewcommand{\arraystretch}{1.1}

\begin{center}
\begin{longtable}{r l r r r r r}
\caption{Human percentile baselines across 50 COMPASS tasks.}\label{tab:human-baselines}\\
\toprule
\textbf{Task Number} & \textbf{Task Name} & \textbf{Mean (\%)} & \textbf{Median (\%)} & \textbf{SD} & \textbf{Skew} & \textbf{Submissions} \\
\midrule
\endfirsthead
\multicolumn{7}{c}{\small\itshape Table \thetable{} (continued)}\\
\toprule
\textbf{Task Number} & \textbf{Task Name} & \textbf{Mean (\%)} & \textbf{Median (\%)} & \textbf{SD} & \textbf{Skew} & \textbf{Submissions} \\
\midrule
\endhead
\midrule
\multicolumn{7}{r}{\small\itshape Continued on next page}
\endfoot
\bottomrule
\endlastfoot
1  & array\_closest\_ascenders            & 25.6 & 0   & 35.4 &  2.17 & 6,001 \\
2  & ascending\_paths                    & 19.2 & 0   & 32.0 &  1.80 & 4,143 \\
3  & balanced\_password                  & 23.1 & 9   & 28.4 &  1.49 & 3,640 \\
4  & ball\_switch\_board                 & 30.8 & 20  & 33.6 &  0.96 & 4,923 \\
5  & beautiful\_password                 & 25.9 & 12  & 32.0 &  1.30 & 2,759 \\
6  & boat\_alignment                     & 22.4 & 0   & 33.2 &  2.03 & 6,321 \\
7  & brackets\_rotation                  & 19.6 & 0   & 29.2 &  2.01 & 6,303 \\
8  & cannonballs                         & 24.3 & 0   & 35.1 &  2.08 & 8,079 \\
9  & cartesian\_sequence                 & 13.3 & 0   & 28.7 &  1.39 & 4,436 \\
10 & clocks                              & 23.9 & 14  & 31.5 &  0.94 & 8,160 \\
11 & count\_bounded\_slices              & 37.0 & 40  & 33.2 & -0.27 & 27,879 \\
12 & count\_palindromic\_slices          & 26.3 & 15.38 & 29.7 &  1.11 & 12,379 \\
13 & different\_characters               & 32.8 & 28  & 33.9 &  0.43 & 5,310 \\
14 & double\_median                      & 13.0 & 0   & 26.9 &  1.45 & 2,462 \\
15 & even\_sums\_game                    & 12.9 & 0   & 27.4 &  1.41 & 7,441 \\
16 & fill\_the\_gaps                     & 21.3 & 0   & 32.9 &  1.95 & 2,614 \\
17 & flooded\_island                     & 40.4 & 44  & 35.7 & -0.30 & 4,948 \\
18 & grocery\_store                      & 22.7 & 0   & 34.1 &  2.00 & 8,974 \\
19 & hamiltonian\_routes\_count          & 11.0 & 0   & 27.0 &  1.23 & 3,084 \\
20 & hit\_the\_number                    & 20.5 & 0   & 29.0 &  2.12 & 5,955 \\
21 & increasing\_sequences               & 20.3 & 11  & 26.8 &  1.04 & 5,464 \\
22 & leader\_slice\_inc                  & 22.1 & 0   & 30.8 &  2.15 & 10,312 \\
23 & letter\_cover                       & 13.6 & 0   & 24.0 &  1.71 & 30,257 \\
24 & longest\_nonnegative\_sum\_slice    & 35.0 & 36  & 32.1 & -0.09 & 5,837 \\
25 & max\_distance\_monotonic            & 33.3 & 22  & 34.8 &  0.98 & 8,639 \\
26 & max\_not\_present                   & 22.2 & 6   & 30.6 &  1.59 & 5,419 \\
27 & max\_path\_from\_the\_left\_top\_corner & 36.5 & 44  & 35.6 & -0.63 & 8,313 \\
28 & max\_square\_on\_matrix             & 30.3 & 15  & 33.5 &  1.37 & 6,036 \\
29 & max\_zero\_product                  & 20.3 & 0   & 30.0 &  2.03 & 5,305 \\
30 & min\_abs\_sum                       & 35.0 & 27  & 34.9 &  0.69 & 74,229 \\
31 & min\_router\_peripherality          & 16.6 & 0   & 29.0 &  1.72 & 4,146 \\
32 & min\_trailing\_zeros                & 26.2 & 5   & 33.9 &  1.88 & 4,887 \\
33 & minfuds                             & 15.2 & 0   & 29.0 &  1.57 & 3,959 \\
34 & multivitamin                        & 27.5 & 23  & 29.4 &  0.46 & 8,849 \\
35 & number\_of\_zeros                   & 16.1 & 0   & 26.9 &  1.80 & 4,196 \\
36 & odd\_network                        & 18.1 & 0   & 31.4 &  1.73 & 3,493 \\
37 & palindromes                         & 17.4 & 0   & 27.5 &  1.90 & 2,694 \\
38 & pets\_and\_toys                     & 20.3 & 0   & 33.6 &  1.81 & 4,514 \\
39 & prefix\_max\_product                & 36.0 & 37  & 30.1 & -0.10 & 7,278 \\
40 & public\_transport\_tickets\_algo    & 20.1 & 0   & 31.1 &  1.94 & 6,081 \\
41 & refueling                           & 20.0 & 0   & 30.8 &  1.95 & 3,818 \\
42 & replacing\_books                    & 29.5 & 7   & 34.3 &  1.96 & 5,496 \\
43 & sheep\_and\_sunshades               & 34.4 & 35  & 32.8 & -0.06 & 6,689 \\
44 & sprinklers\_arrangement             & 24.4 & 0   & 35.8 &  2.04 & 2,967 \\
45 & stones                              & 12.3 & 0   & 26.1 &  1.42 & 2,717 \\
46 & string\_modification                & 21.7 & 0   & 31.9 &  2.04 & 3,158 \\
47 & theater\_tickets                    & 28.0 & 17  & 30.9 &  1.07 & 4,441 \\
48 & three\_letters\_blocks              & 12.9 & 0   & 26.2 &  1.47 & 3,478 \\
49 & trek\_and\_swim                     & 14.3 & 0   & 30.3 &  1.41 & 8,334 \\
50 & trip\_planning                      & 20.0 & 0   & 31.1 &  1.94 & 6,333 \\
\end{longtable}
\end{center}
\clearpage             

\section{Correctness Scores by Task and Model}
The table below reports the median, mean, and standard deviation of correctness scores for each model across every task. A score of 100 indicates that the model’s submitted solution was fully correct and accepted by all test cases. Lower scores reflect partially correct or failed solutions.
\noindent\emph{Abbreviations: C = \texttt{Claude-Opus-4}, 
G = \texttt{Gemini-2.5-Pro}, 
O = \texttt{O4-Mini-High}}

\scriptsize
\setlength{\tabcolsep}{5pt}
\renewcommand{\arraystretch}{1.1}

\begin{longtable}{r r r r r r r r r r}
\caption{Correctness by task and model: medians, means, and standard deviations.}
\label{tab:correctness-by-task}\\
\toprule
\textbf{Task Number} &
\textbf{C Med} & \textbf{G Med} & \textbf{O Med} &
\textbf{C Mean} & \textbf{G Mean} & \textbf{O Mean} &
\textbf{C SD} & \textbf{G SD} & \textbf{O SD} \\
\midrule
\endfirsthead
\multicolumn{10}{c}{\small\itshape Table \thetable{} (continued)}\\
\toprule
\textbf{Task Number} &
\textbf{C Med} & \textbf{G Med} & \textbf{O Med} &
\textbf{C Mean} & \textbf{G Mean} & \textbf{O Mean} &
\textbf{C SD} & \textbf{G SD} & \textbf{O SD} \\
\midrule
\endhead
\midrule
\multicolumn{10}{r}{\small\itshape Continued on next page}
\endfoot
\bottomrule
\endlastfoot

 1 & 100.0 & 100.0 & 100.0 &  82.5 & 100.0 &  98.4 & 33.3 &  0.0 & 12.5 \\
 2 &  40.0 & 100.0 & 100.0 &  43.1 & 100.0 & 100.0 & 20.8 &  0.0 &  0.0 \\
 3 & 100.0 & 100.0 & 100.0 &  95.8 & 100.0 &  97.1 & 19.2 &  0.0 & 14.4 \\
 4 & 100.0 & 100.0 & 100.0 &  76.2 &  94.7 &  99.1 & 28.0 & 21.2 &  7.5 \\
 5 & 100.0 & 100.0 & 100.0 &  89.1 & 100.0 & 100.0 & 31.5 &  0.0 &  0.0 \\
 6 & 100.0 & 100.0 & 100.0 &  87.5 &  98.4 & 100.0 & 24.6 & 12.5 &  0.0 \\
 7 &  25.0 &  58.3 &  58.3 &  32.9 &  61.1 &  57.2 & 23.0 & 25.4 &  4.2 \\
 8 & 100.0 & 100.0 & 100.0 &  99.4 & 100.0 &  98.4 &  3.5 &  0.0 & 12.5 \\
 9 &   0.0 & 100.0 & 100.0 &   4.2 &  94.2 & 100.0 &  7.5 & 22.7 &  0.0 \\
10 &  37.5 & 100.0 & 100.0 &  46.1 &  98.0 &  98.4 & 15.2 & 11.0 & 12.5 \\
11 & 100.0 & 100.0 & 100.0 & 100.0 & 100.0 & 100.0 &  0.0 &  0.0 &  0.0 \\
12 & 100.0 & 100.0 & 100.0 &  59.4 & 100.0 & 100.0 & 49.5 &  0.0 &  0.0 \\
13 &  66.7 & 100.0 & 100.0 &  66.5 &  95.7 &  94.4 &  6.4 & 10.4 & 21.6 \\
14 &  42.9 & 100.0 & 100.0 &  47.1 & 100.0 &  96.0 & 46.6 &  0.0 & 18.8 \\
15 &   0.0 &  28.6 & 100.0 &  11.8 &  28.1 & 100.0 & 25.9 & 26.9 &  0.0 \\
16 &  14.3 & 100.0 & 100.0 &  15.0 &  82.6 &  87.5 & 17.2 & 33.7 & 28.3 \\
17 & 100.0 & 100.0 & 100.0 & 100.0 & 100.0 & 100.0 &  0.0 &  0.0 &  0.0 \\
18 & 100.0 & 100.0 & 100.0 & 100.0 & 100.0 & 100.0 &  0.0 &  0.0 &  0.0 \\
19 &  50.0 & 100.0 & 100.0 &  71.1 & 100.0 & 100.0 & 24.9 &  0.0 &  0.0 \\
20 &  83.3 & 100.0 & 100.0 &  68.2 &  96.6 &  94.8 & 32.1 & 17.6 & 20.8 \\
21 & 100.0 & 100.0 & 100.0 & 100.0 & 100.0 & 100.0 &  0.0 &  0.0 &  0.0 \\
22 &  80.0 & 100.0 & 100.0 &  85.6 &  91.6 &  87.5 & 17.6 & 25.1 & 28.2 \\
23 & 100.0 & 100.0 & 100.0 & 100.0 & 100.0 & 100.0 &  0.0 &  0.0 &  0.0 \\
24 &  66.7 & 100.0 & 100.0 &  55.7 & 100.0 & 100.0 & 44.4 &  0.0 &  0.0 \\
25 & 100.0 & 100.0 & 100.0 &  96.6 &  96.2 & 100.0 & 10.4 & 18.1 &  0.0 \\
26 &   0.0 & 100.0 & 100.0 &  10.7 &  67.7 & 100.0 & 17.7 & 41.6 &  0.0 \\
27 & 100.0 & 100.0 & 100.0 & 100.0 &  97.7 & 100.0 &  0.0 & 11.1 &  0.0 \\
28 & 100.0 & 100.0 & 100.0 &  92.2 & 100.0 & 100.0 & 27.0 &  0.0 &  0.0 \\
29 & 100.0 & 100.0 & 100.0 &  98.8 & 100.0 &  97.5 & 10.0 &  0.0 & 13.1 \\
30 & 100.0 & 100.0 & 100.0 &  90.9 &  96.9 & 100.0 & 28.6 & 17.5 &  0.0 \\
31 & 100.0 & 100.0 & 100.0 & 100.0 &  98.4 & 100.0 &  0.0 & 12.5 &  0.0 \\
32 & 100.0 & 100.0 & 100.0 &  93.4 &  97.9 & 100.0 &  8.6 &  6.5 &  0.0 \\
33 & 100.0 & 100.0 & 100.0 & 100.0 & 100.0 &  99.0 &  0.0 &  0.0 &  5.8 \\
34 & 100.0 & 100.0 & 100.0 &  99.4 &  99.8 &  99.6 &  4.7 &  1.6 &  3.1 \\
35 &  50.0 &  75.0 & 100.0 &  55.9 &  83.2 & 100.0 & 26.6 & 19.4 &  0.0 \\
36 & 100.0 & 100.0 & 100.0 & 100.0 & 100.0 & 100.0 &  0.0 &  0.0 &  0.0 \\
37 &  33.3 & 100.0 &  75.0 &  35.9 & 100.0 &  66.7 & 17.1 &  0.0 & 34.2 \\
38 & 100.0 & 100.0 & 100.0 & 100.0 & 100.0 & 100.0 &  0.0 &  0.0 &  0.0 \\
39 & 100.0 & 100.0 & 100.0 &  98.0 & 100.0 &  98.4 & 11.9 &  0.0 & 12.5 \\
40 & 100.0 & 100.0 & 100.0 & 100.0 & 100.0 & 100.0 &  0.0 &  0.0 &  0.0 \\
41 &  80.0 & 100.0 & 100.0 &  75.6 &  97.5 &  97.2 & 12.7 & 12.0 & 13.9 \\
42 &  66.7 & 100.0 & 100.0 &  70.7 &  98.3 & 100.0 & 13.5 & 12.6 &  0.0 \\
43 & 100.0 & 100.0 & 100.0 &  93.2 &  98.2 &  68.8 & 17.0 & 12.6 & 46.7 \\
44 &  40.0 & 100.0 & 100.0 &  47.3 & 100.0 &  98.4 & 36.2 &  0.0 & 12.5 \\
45 &  28.6 & 100.0 &  57.1 &  29.0 &  78.8 &  57.8 &  2.5 & 29.7 & 40.8 \\
46 &  28.6 & 100.0 & 100.0 &  33.3 &  80.1 &  98.0 & 21.6 & 33.1 &  5.6 \\
47 &  33.3 & 100.0 & 100.0 &  55.9 &  96.4 &  98.8 & 28.8 & 14.3 &  7.4 \\
48 &  91.7 & 100.0 & 100.0 &  68.1 &  94.1 &  97.3 & 37.5 & 23.0 & 15.5 \\
49 & 100.0 & 100.0 & 100.0 &  92.2 &  93.8 &  95.9 & 22.7 & 19.8 & 17.5 \\
50 &   0.0 & 100.0 & 100.0 &  34.6 &  75.2 &  97.8 & 41.9 & 36.0 & 11.4 \\
\end{longtable}

\clearpage
\section{Efficiency Scores by Task and Model}

The table below reports the median, mean, and standard deviation of efficiency scores for each model across every task. A score of 100 indicates the solution was highly optimized, with minimal runtime, memory usage, and/or number of attempts. Lower scores indicate less efficient performance.
\noindent\emph{Abbreviations: C = \texttt{Claude-Opus-4},\;
G = \texttt{Gemini-2.5-Pro},\;
O = \texttt{O4-Mini-High}.}

\begingroup
\scriptsize
\setlength{\tabcolsep}{5pt}
\renewcommand{\arraystretch}{1.1}

\begin{longtable}{r r r r r r r r r r}
\caption{Efficiency by task and model: medians, means, and standard deviations.}
\label{tab:appC-efficiency}\\
\toprule
\textbf{Task Number} &
\textbf{C Med} & \textbf{G Med} & \textbf{O Med} &
\textbf{C Mean} & \textbf{G Mean} & \textbf{O Mean} &
\textbf{C SD} & \textbf{G SD} & \textbf{O SD} \\
\midrule
\endfirsthead
\multicolumn{10}{c}{\small\itshape Table \thetable{} (continued)}\\
\toprule
\textbf{Task Number} &
\textbf{C Med} & \textbf{G Med} & \textbf{O Med} &
\textbf{C Mean} & \textbf{G Mean} & \textbf{O Mean} &
\textbf{C SD} & \textbf{G SD} & \textbf{O SD} \\
\midrule
\endhead
\midrule
\multicolumn{10}{r}{\small\itshape Continued on next page}
\endfoot
\bottomrule
\endlastfoot

 1  & 100.0 & 100.0 & 100.0 & 62.5 & 100.0 & 98.4 & 48.8 & 0.0 & 12.5 \\
 2  &  25.0 & 100.0 & 100.0 & 27.7 & 100.0 & 100.0 & 26.0 & 0.0 & 0.0 \\
 3  &   0.0 &  20.0 &  60.0 &  0.0 &  19.7 &  55.0 &  0.0 & 2.5 & 25.2 \\
 4  &  40.0 &  40.0 & 100.0 & 25.9 &  53.1 &  98.4 & 18.1 & 28.6 & 12.5 \\
 5  & 100.0 & 100.0 & 100.0 & 88.3 & 100.0 & 100.0 & 31.8 & 0.0 & 0.0 \\
 6  &  57.1 &  57.1 &  85.7 & 41.3 &  65.0 &  89.7 & 29.8 & 13.9 & 10.0 \\
 7  &   0.0 &   0.0 &  33.3 &  1.0 &  15.1 &  29.7 &  5.8 & 16.7 & 13.4 \\
 8  &   0.0 & 100.0 & 100.0 &  1.3 & 100.0 &  98.4 &  8.6 & 0.0 & 12.5 \\
 9  &   0.0 &  50.0 & 100.0 &  0.3 &  62.2 &  93.2 &  2.1 & 35.4 & 18.5 \\
10  &   0.0 & 100.0 & 100.0 &  9.1 &  94.6 &  98.4 & 21.4 & 18.7 & 12.5 \\
11  & 100.0 & 100.0 & 100.0 & 93.4 & 100.0 & 100.0 & 20.8 &  0.0 &  0.0 \\
12  &  33.3 & 100.0 & 100.0 & 20.5 &  98.3 & 100.0 & 17.3 &  9.9 &  0.0 \\
13  &  33.3 & 100.0 & 100.0 & 24.5 &  90.6 &  93.8 & 18.1 & 21.8 & 24.4 \\
14  &  18.2 & 100.0 & 100.0 & 49.9 & 100.0 &  96.7 & 39.8 &  0.0 & 17.5 \\
15  &   0.0 &   0.0 & 100.0 &  0.0 &  15.6 & 100.0 &  0.0 & 27.6 &  0.0 \\
16  &   0.0 & 100.0 & 100.0 &  2.1 &  70.3 &  81.8 & 13.1 & 42.9 & 34.1 \\
17  &   0.0 & 100.0 & 100.0 &  0.8 & 100.0 & 100.0 &  4.4 &  0.0 &  0.0 \\
18  & 100.0 & 100.0 & 100.0 & 100.0 & 100.0 & 100.0 &  0.0 &  0.0 &  0.0 \\
19  &   0.0 & 100.0 & 100.0 &  0.0 & 100.0 & 100.0 &  0.0 &  0.0 &  0.0 \\
20  &  22.2 & 100.0 & 100.0 & 43.4 &  92.4 &  94.4 & 42.4 & 26.0 & 21.9 \\
21  & 100.0 & 100.0 & 100.0 & 100.0 & 100.0 & 100.0 &  0.0 &  0.0 &  0.0 \\
22  &  12.5 & 100.0 & 100.0 & 13.7 &  92.6 &  94.1 & 15.4 & 23.4 & 17.7 \\
23  & 100.0 & 100.0 & 100.0 & 100.0 & 100.0 & 100.0 &  0.0 &  0.0 &  0.0 \\
24  &  33.3 & 100.0 & 100.0 & 33.1 & 100.0 & 100.0 & 23.8 &  0.0 &  0.0 \\
25  & 100.0 & 100.0 & 100.0 & 77.0 &  94.5 & 100.0 & 40.2 & 22.0 &  0.0 \\
26  &   0.0 &  60.0 & 100.0 &  5.9 &  51.6 & 100.0 & 13.7 & 38.7 &  0.0 \\
27  &  33.3 &  33.3 & 100.0 & 32.8 &  51.6 &  99.0 &  4.2 & 38.0 &  8.3 \\
28  &   0.0 & 100.0 & 100.0 & 16.1 & 100.0 & 100.0 & 17.8 &  0.0 &  0.0 \\
29  &  25.0 & 100.0 & 100.0 & 34.8 & 100.0 &  98.4 & 38.5 &  0.0 & 12.5 \\
30  &  40.0 & 100.0 & 100.0 & 31.6 &  96.9 & 100.0 & 13.2 & 17.5 &  0.0 \\
31  &  14.3 & 100.0 & 100.0 & 34.4 &  98.4 &  99.8 & 32.4 & 12.5 &  1.8 \\
32  & 100.0 & 100.0 & 100.0 & 99.3 &  99.0 & 100.0 &  2.7 &  3.3 &  0.0 \\
33  &  40.0 & 100.0 & 100.0 & 37.5 &  95.6 &  98.8 &  6.7 & 11.0 & 10.0 \\
34  &   0.0 & 100.0 & 100.0 &  5.3 &  99.1 &  99.1 & 12.5 &  7.5 &  7.5 \\
35  &   0.0 & 100.0 & 100.0 & 10.9 &  95.5 & 100.0 & 26.9 & 20.5 &  0.0 \\
36  & 100.0 & 100.0 & 100.0 & 100.0 & 100.0 & 100.0 &  0.0 &  0.0 &  0.0 \\
37  &  20.0 & 100.0 &  50.0 & 20.0 & 100.0 &  58.8 &  0.0 &  0.0 & 38.5 \\
38  & 100.0 & 100.0 & 100.0 & 83.6 & 100.0 & 100.0 & 23.7 &  0.0 &  0.0 \\
39  &  22.2 & 100.0 & 100.0 & 25.5 & 100.0 &  98.4 & 11.1 &  0.0 & 12.5 \\
40  & 100.0 & 100.0 & 100.0 & 99.2 & 100.0 & 100.0 &  6.3 &  0.0 &  0.0 \\
41  &  33.3 & 100.0 & 100.0 & 32.0 &  96.6 &  96.6 & 12.4 & 16.0 & 16.0 \\
42  &  37.5 & 100.0 & 100.0 & 47.1 &  98.2 & 100.0 & 22.2 & 12.6 &  0.0 \\
43  &   0.0 & 100.0 & 100.0 &  4.7 &  98.2 &  68.8 & 11.3 & 12.6 & 46.7 \\
44  &   0.0 & 100.0 & 100.0 & 29.3 & 100.0 &  98.4 & 44.9 &  0.0 & 12.5 \\
45  &   0.0 & 100.0 &  40.0 &  0.6 &  70.6 &  51.6 &  3.5 & 39.0 & 42.7 \\
46  &  25.0 & 100.0 & 100.0 & 23.8 &  82.8 &  99.2 &  5.3 & 34.8 &  6.2 \\
47  &  12.5 & 100.0 & 100.0 & 15.4 &  94.9 &  98.8 & 15.2 & 20.0 &  9.4 \\
48  &  60.0 &  70.0 & 100.0 & 43.1 &  73.6 &  90.8 & 29.8 & 22.1 & 19.1 \\
49  &  20.0 &  40.0 &  80.0 & 23.1 &  59.1 &  77.5 & 13.0 & 30.9 & 23.2 \\
50  &   0.0 &  16.7 & 100.0 &  0.0 &  42.4 &  96.1 &  0.0 & 38.9 & 16.7 \\

\end{longtable}
\endgroup

\clearpage
\section{Quality Scores by Task and Model}

The table below reports the median, mean, and standard deviation of quality scores for each model across every task. A score of 100 reflects the highest quality solutions, based on criteria such as code readability, structure, naming, and adherence to clean coding practices. Lower scores indicate suboptimal or poorly structured code.
\noindent\emph{Abbreviations: C = \texttt{Claude-Opus-4},\;
G = \texttt{Gemini-2.5-Pro},\;
O = \texttt{O4-Mini-High}.}

\begingroup
\scriptsize
\setlength{\tabcolsep}{5pt}
\renewcommand{\arraystretch}{1.1}

\begin{longtable}{r r r r r r r r r r}
\caption{Quality by task and model: medians, means, and standard deviations.}
\label{tab:appD-quality}\\
\toprule
\textbf{Task Number} &
\textbf{C Med} & \textbf{G Med} & \textbf{O Med} &
\textbf{C Mean} & \textbf{G Mean} & \textbf{O Mean} &
\textbf{C SD} & \textbf{G SD} & \textbf{O SD} \\
\midrule
\endfirsthead
\multicolumn{10}{c}{\small\itshape Table \thetable{} (continued)}\\
\toprule
\textbf{Task Number} &
\textbf{C Med} & \textbf{G Med} & \textbf{O Med} &
\textbf{C Mean} & \textbf{G Mean} & \textbf{O Mean} &
\textbf{C SD} & \textbf{G SD} & \textbf{O SD} \\
\midrule
\endhead
\midrule
\multicolumn{10}{r}{\small\itshape Continued on next page}
\endfoot
\bottomrule
\endlastfoot

 1  & 95.1 & 94.6 & 95.3 & 94.5 & 94.7 & 95.2 & 2.1 & 0.5 & 0.8 \\
 2  & 92.4 & 96.8 & 89.5 & 92.6 & 95.0 & 91.0 & 4.5 & 4.8 & 2.5 \\
 3  & 92.4 & 89.3 & 86.7 & 92.3 & 89.3 & 85.7 & 1.5 & 3.0 & 2.9 \\
 4  & 88.2 & 88.8 & 89.5 & 89.2 & 88.9 & 91.1 & 4.0 & 4.1 & 3.2 \\
 5  & 100.0 & 100.0 & 98.4 & 99.8 & 99.8 & 98.6 & 0.7 & 0.6 & 0.5 \\
 6  & 95.3 & 98.4 & 95.3 & 95.0 & 96.7 & 95.2 & 4.5 & 2.0 & 1.8 \\
 7  & 89.5 & 89.5 & 94.0 & 92.2 & 90.8 & 92.6 & 5.3 & 2.3 & 3.4 \\
 8  & 98.4 & 94.4 & 92.4 & 98.1 & 94.3 & 91.7 & 1.2 & 1.9 & 3.2 \\
 9  & 91.4 & 93.5 & 94.5 & 90.7 & 92.7 & 92.4 & 3.4 & 3.2 & 3.8 \\
10  & 96.0 & 93.4 & 92.4 & 96.9 & 92.9 & 91.0 & 2.4 & 2.6 & 2.8 \\
11  & 92.4 & 100.0 & 100.0 & 93.0 & 99.7 & 98.7 & 1.8 & 1.1 & 2.9 \\
12  & 92.4 & 100.0 & 89.5 & 90.9 & 98.5 & 91.2 & 3.4 & 2.5 & 3.3 \\
13  & 90.9 & 94.8 & 88.5 & 91.9 & 93.2 & 89.0 & 3.7 & 3.6 & 4.6 \\
14  & 96.8 & 96.8 & 95.3 & 94.8 & 95.5 & 93.7 & 2.8 & 2.1 & 3.8 \\
15  & 88.5 & 94.5 & 42.0 & 87.8 & 92.7 & 45.9 & 3.0 & 4.9 & 10.0 \\
16  & 78.5 & 94.0 & 86.0 & 79.0 & 91.5 & 85.4 & 4.2 & 5.0 & 6.3 \\
17  & 93.8 & 92.2 & 90.4 & 93.7 & 93.1 & 92.4 & 0.8 & 4.7 & 4.5 \\
18  & 95.3 & 94.8 & 95.3 & 94.3 & 94.0 & 95.4 & 1.6 & 2.0 & 1.5 \\
19  & 85.7 & 83.4 & 38.1 & 84.0 & 81.6 & 41.5 & 3.3 & 6.9 & 7.3 \\
20  & 89.5 & 94.3 & 92.4 & 89.1 & 92.2 & 91.2 & 3.9 & 3.4 & 3.3 \\
21  & 96.7 & 96.6 & 96.8 & 96.6 & 96.6 & 98.0 & 0.5 & 0.9 & 1.6 \\
22  & 95.3 & 94.5 & 94.0 & 95.9 & 93.8 & 92.1 & 2.8 & 3.7 & 3.6 \\
23  & 95.0 & 82.8 & 47.2 & 94.2 & 84.0 & 47.8 & 3.6 & 6.5 & 7.8 \\
24  & 96.8 & 95.3 & 95.3 & 96.2 & 96.4 & 96.5 & 3.8 & 2.2 & 1.9 \\
25  & 98.4 & 95.3 & 95.3 & 97.5 & 95.6 & 96.5 & 3.0 & 1.6 & 1.6 \\
26  & 88.9 & 87.5 & 49.0 & 90.0 & 88.6 & 47.7 & 6.7 & 6.5 & 8.9 \\
27  & 96.8 & 92.7 & 89.5 & 95.6 & 91.9 & 90.8 & 2.7 & 5.5 & 3.3 \\
28  & 100.0 & 95.3 & 95.3 & 97.6 & 95.3 & 95.6 & 3.0 & 1.8 & 1.7 \\
29  & 89.0 & 89.2 & 87.9 & 91.5 & 89.1 & 86.7 & 5.6 & 3.8 & 3.1 \\
30  & 95.3 & 89.0 & 98.4 & 94.7 & 89.3 & 97.4 & 4.9 & 3.4 & 2.0 \\
31  & 95.3 & 89.9 & 88.1 & 93.6 & 90.9 & 87.9 & 3.0 & 2.6 & 3.6 \\
32  & 80.8 & 92.4 & 87.4 & 81.0 & 90.8 & 86.9 & 4.7 & 4.3 & 2.7 \\
33  & 96.8 & 98.4 & 87.2 & 96.6 & 98.2 & 87.0 & 3.1 & 1.5 & 4.9 \\
34  & 98.4 & 94.8 & 96.4 & 98.7 & 93.7 & 95.3 & 0.9 & 3.0 & 2.9 \\
35  & 94.8 & 100.0 & 73.8 & 91.5 & 97.6 & 82.6 & 5.5 & 3.3 & 16.5 \\
36  & 96.8 & 100.0 & 93.8 & 97.4 & 96.6 & 94.8 & 2.5 & 4.0 & 2.8 \\
37  & 84.5 & 83.3 & 78.1 & 86.7 & 84.7 & 79.9 & 6.2 & 5.9 & 8.7 \\
38  & 88.5 & 89.3 & 94.8 & 88.7 & 90.6 & 94.7 & 0.8 & 2.6 & 1.2 \\
39  & 88.5 & 100.0 & 94.5 & 91.1 & 98.8 & 91.5 & 5.3 & 1.8 & 4.7 \\
40  & 96.6 & 100.0 & 100.0 & 96.9 & 99.9 & 99.5 & 1.4 & 0.5 & 1.2 \\
41  & 92.4 & 88.0 & 87.9 & 92.3 & 88.1 & 86.6 & 4.3 & 3.7 & 3.0 \\
42  & 92.4 & 100.0 & 100.0 & 93.6 & 98.4 & 99.7 & 2.8 & 2.3 & 1.2 \\
43  & 100.0 & 91.4 & 87.6 & 99.5 & 92.3 & 86.9 & 2.1 & 3.5 & 3.4 \\
44  & 82.2 & 100.0 & 100.0 & 85.2 & 100.0 & 99.9 & 8.4 & 0.0 & 0.4 \\
45  & 100.0 & 100.0 & 96.8 & 94.0 & 99.7 & 95.6 & 7.2 & 1.7 & 4.3 \\
46  & 80.0 & 92.6 & 92.8 & 79.9 & 89.6 & 91.4 & 4.5 & 5.6 & 3.4 \\
47  & 95.8 & 100.0 & 96.8 & 95.0 & 99.9 & 96.4 & 5.4 & 0.6 & 3.7 \\
48  & 87.5 & 87.3 & 93.5 & 86.3 & 87.5 & 91.7 & 5.2 & 5.7 & 5.0 \\
49  & 93.8 & 92.2 & 87.9 & 92.5 & 91.6 & 86.9 & 2.1 & 3.6 & 4.2 \\
50  & 86.7 & 78.8 & 82.4 & 86.5 & 78.6 & 82.5 & 1.5 & 4.6 & 0.6 \\
\end{longtable}
\endgroup

\clearpage
\section{Composite Scores by Task and Model}

The table below reports the median, mean, and standard deviation of composite scores for each model across every task. These scores combine correctness, efficiency, and code quality into a single metric, offering a holistic view of model performance. A score of 100 represents a solution that is fully correct, highly efficient, and of excellent quality.
\noindent\emph{Abbreviations: C = \texttt{Claude-Opus-4},\;
G = \texttt{Gemini-2.5-Pro},\;
O = \texttt{O4-Mini-High}.}

\begingroup
\scriptsize
\setlength{\tabcolsep}{5pt}
\renewcommand{\arraystretch}{1.1}

\begin{longtable}{r r r r r r r r r r}
\caption{Composite by task and model: medians, means, and standard deviations.}
\label{tab:app-composite}\\
\toprule
\textbf{Task Number} &
\textbf{C Med} & \textbf{G Med} & \textbf{O Med} &
\textbf{C Mean} & \textbf{G Mean} & \textbf{O Mean} &
\textbf{C SD} & \textbf{G SD} & \textbf{O SD} \\
\midrule
\endfirsthead
\multicolumn{10}{c}{\small\itshape Table \thetable{} (continued)}\\
\toprule
\textbf{Task Number} &
\textbf{C Med} & \textbf{G Med} & \textbf{O Med} &
\textbf{C Mean} & \textbf{G Mean} & \textbf{O Mean} &
\textbf{C SD} & \textbf{G SD} & \textbf{O SD} \\
\midrule
\endhead
\midrule
\multicolumn{10}{r}{\small\itshape Continued on next page}
\endfoot
\bottomrule
\endlastfoot

 1  & 98.2 & 98.2 & 98.4 & 79.8 & 98.2 & 96.9 & 25.1 & 0.2 & 12.3 \\
 2  & 51.5 & 98.9 & 96.5 & 54.5 & 98.3 & 97.0 & 14.6 & 1.6 & 0.8 \\
 3  & 64.1 & 69.8 & 80.4 & 62.7 & 69.7 & 78.8 &  6.8 & 1.3 & 13.0 \\
 4  & 74.8 & 76.6 & 96.5 & 63.8 & 78.4 & 96.2 & 15.3 &16.1 &  6.7 \\
 5  &100.0 &100.0 & 99.5 & 89.3 & 99.9 & 99.5 & 30.2 & 0.2 &  0.2 \\
 6  & 84.1 & 85.2 & 93.7 & 74.6 & 86.2 & 95.0 & 17.5 &11.6 &  3.4 \\
 7  & 39.0 & 57.6 & 61.9 & 42.0 & 55.2 & 59.8 &  8.0 &12.0 &  6.2 \\
 8  & 66.1 & 98.2 & 97.5 & 66.3 & 98.1 & 96.2 &  1.9 & 0.6 &  8.6 \\
 9  &  0.0 & 81.4 & 98.2 & 10.0 & 82.6 & 95.2 & 16.2 &18.8 &  6.5 \\
10  & 45.8 & 97.5 & 97.5 & 50.7 & 95.2 & 95.9 & 11.6 & 9.2 &  8.6 \\
11  & 97.5 &100.0 &100.0 & 95.5 & 99.9 & 99.6 &  6.7 & 0.4 &  1.0 \\
12  & 72.5 &100.0 & 96.5 & 56.9 & 98.9 & 97.1 & 21.8 & 3.5 &  1.1 \\
13  & 63.0 & 98.2 & 96.0 & 60.9 & 93.2 & 92.4 &  7.8 &11.0 & 14.8 \\
14  & 68.6 & 98.9 & 97.9 & 63.9 & 98.5 & 95.5 & 26.0 & 0.7 & 11.9 \\
15  & 29.7 & 38.6 & 85.4 & 33.2 & 42.6 & 88.7 &  8.8 &22.1 &  9.2 \\
16  & 30.5 & 97.7 & 94.8 & 32.0 & 80.0 & 84.4 & 10.2 &28.9 & 22.6 \\
17  & 64.6 & 97.4 & 96.8 & 64.8 & 97.7 & 97.5 &  1.6 & 1.6 &  1.5 \\
18  & 98.4 & 98.3 & 98.4 & 98.1 & 98.0 & 98.5 &  0.5 & 0.7 &  0.5 \\
19  & 45.7 & 94.5 & 79.4 & 51.7 & 94.0 & 82.3 &  8.3 & 2.4 &  6.2 \\
20  & 63.6 & 98.1 & 97.5 & 66.9 & 93.7 & 93.5 & 21.3 &13.4 & 14.3 \\
21  & 98.9 & 98.9 & 98.9 & 98.9 & 98.9 & 99.4 &  0.2 & 0.3 &  0.5 \\
22  & 66.1 & 98.1 & 97.5 & 65.1 & 91.7 & 91.2 &  6.9 &19.4 & 13.0 \\
23  & 98.3 & 94.3 & 82.4 & 98.1 & 94.7 & 83.2 &  1.2 & 2.2 &  4.0 \\
24  & 63.4 & 98.4 & 98.4 & 61.7 & 98.8 & 98.8 & 21.9 & 0.7 &  0.6 \\
25  & 98.4 & 98.4 & 98.4 & 90.3 & 95.5 & 98.8 & 15.4 &13.1 &  0.5 \\
26  & 32.8 & 82.4 &100.0 & 35.5 & 67.0 & 92.4 &  9.5 &30.3 &  8.9 \\
27  & 76.7 & 75.4 & 96.5 & 76.1 & 80.4 & 96.6 &  1.8 &15.5 &  3.0 \\
28  & 66.7 & 98.4 & 98.4 & 68.6 & 98.4 & 98.5 & 12.1 & 0.6 &  0.6 \\
29  & 70.1 & 96.4 & 96.1 & 75.0 & 96.4 & 95.3 & 12.9 & 1.3 &  8.8 \\
30  & 78.4 & 96.3 & 99.5 & 72.4 & 94.3 & 99.1 & 14.1 &11.9 &  0.7 \\
31  & 69.9 & 96.6 & 96.0 & 76.0 & 95.9 & 95.9 & 11.1 & 8.4 &  1.2 \\
32  & 91.3 & 97.4 & 95.8 & 91.2 & 95.9 & 95.6 &  3.6 & 3.4 &  0.9 \\
33  & 78.9 & 99.5 & 95.6 & 78.0 & 97.9 & 94.9 &  3.0 & 3.5 &  5.7 \\
34  & 66.1 & 98.2 & 98.8 & 67.8 & 97.5 & 98.0 &  3.7 & 2.9 &  3.5 \\
35  & 48.4 & 91.7 & 91.3 & 52.8 & 92.1 & 94.2 & 16.4 &12.2 &  5.5 \\
36  & 98.9 &100.0 & 97.9 & 99.1 & 98.9 & 98.3 &  0.8 & 1.3 &  0.9 \\
37  & 49.0 & 94.8 & 69.5 & 47.6 & 95.8 & 67.6 &  5.8 & 2.6 & 24.4 \\
38  & 96.1 & 96.4 & 98.3 & 90.8 & 96.9 & 98.2 &  7.8 & 0.9 &  0.4 \\
39  & 70.5 &100.0 & 98.2 & 71.5 & 99.6 & 95.7 &  3.9 & 0.6 & 12.2 \\
40  & 98.9 &100.0 &100.0 & 98.7 &100.0 & 99.8 &  2.1 & 0.2 &  0.4 \\
41  & 68.6 & 96.0 & 95.8 & 66.7 & 94.1 & 93.5 &  7.5 & 9.8 &  9.7 \\
42  & 65.5 &100.0 &100.0 & 70.4 & 98.3 & 99.9 & 12.2 & 8.3 &  0.4 \\
43  & 66.7 & 96.5 & 94.6 & 65.8 & 95.7 & 65.8 &  4.9 &12.3 & 44.7 \\
44  & 41.6 &100.0 &100.0 & 53.1 &100.0 & 98.4 & 29.2 & 0.0 & 12.5 \\
45  & 42.9 &100.0 & 65.7 & 41.2 & 82.0 & 59.8 &  3.1 &25.7 & 39.9 \\
46  & 45.3 & 94.0 & 97.4 & 45.7 & 81.4 & 96.2 &  8.1 &29.8 &  3.8 \\
47  & 47.8 &100.0 & 98.9 & 55.4 & 97.1 & 98.0 & 15.1 &11.4 &  5.5 \\
48  & 80.4 & 85.8 & 97.6 & 65.8 & 84.2 & 93.3 & 22.8 &18.7 & 11.3 \\
49  & 71.3 & 78.3 & 89.3 & 69.3 & 81.5 & 86.8 & 10.8 &14.8 & 13.1 \\
50  & 29.5 & 66.7 & 94.1 & 40.4 & 62.2 & 92.1 & 13.9 &28.4 &  9.0 \\
\end{longtable}
\endgroup

\clearpage
\section{Combined Correctness and Efficiency Percentile Scores by Task and Model}

The table below shows the combined correctness and efficiency scores for each model, alongside the human average performance for each task. For each model, the raw score is followed by its percentile rank in parentheses, calculated relative to human performance. This allows for direct, task-by-task comparisons between AI-generated code and competitive human submissions. The percentile scores highlight not just how well each model performs, but how often that performance would outperform human programmers under similar constraints.

\begingroup
\scriptsize
\setlength{\tabcolsep}{5pt}
\renewcommand{\arraystretch}{1.1}

\begin{longtable}{r l r r r r}
\caption{Combined Correctness and Efficiency: Human vs.\ AI (raw score with human-relative percentile in parentheses).}
\label{tab:appA6-percentiles}\\
\toprule
\textbf{Task \#} & \textbf{Difficulty} & \textbf{Human Mean} & \textbf{Claude-Opus-4} & \textbf{Gemini-2.5-Pro} & \textbf{O4-Mini-High} \\
\midrule
\endfirsthead
\multicolumn{6}{c}{\small\itshape Table \thetable{} (continued)}\\
\toprule
\textbf{Task \#} & \textbf{Difficulty} & \textbf{Human Mean} & \textbf{C} & \textbf{G} & \textbf{O} \\
\midrule
\endhead
\midrule
\multicolumn{6}{r}{\small\itshape Continued on next page}
\endfoot
\bottomrule
\endlastfoot

 1  & medium & 25.6 & 72.5 (91) & 100.0 (98) & 98.4 (98) \\
 2  & hard   & 19.2 & 35.4 (69) & 100.0 (99) & 100.0 (99) \\
 3  & hard   & 23.1 & 47.9 (81) & 59.9 (90)  & 76.0 (97)  \\
 4  & hard   & 30.8 & 51.0 (73) & 73.9 (90)  & 98.8 (98)  \\
 5  & medium & 25.9 & 88.7 (98) & 100.0 (99) & 100.0 (99) \\
 6  & hard   & 22.4 & 64.4 (90) & 81.7 (96)  & 94.8 (99)  \\
 7  & hard   & 19.6 & 16.9 (46) & 38.1 (74)  & 43.5 (79)  \\
 8  & hard   & 24.3 & 50.4 (77) & 100.0 (98) & 98.4 (98)  \\
 9  & hard   & 13.3 &  2.2 (35) & 78.2 (99)  & 96.6 (100) \\
10  & hard   & 23.9 & 27.6 (55) & 96.3 (99)  & 98.4 (99)  \\
11  & medium & 37.0 & 96.7 (96) & 100.0 (97) & 100.0 (97) \\
12  & hard   & 26.3 & 40.0 (68) & 99.2 (99)  & 100.0 (99) \\
13  & medium & 32.8 & 45.5 (65) & 93.2 (96)  & 94.1 (96)  \\
14  & hard   & 13.0 & 48.5 (91) & 100.0 (100)& 96.3 (100) \\
15  & hard   & 12.9 &  5.9 (40) & 21.9 (63)  & 100.0 (100) \\
16  & hard   & 21.3 &  8.6 (35) & 76.5 (95)  & 84.7 (97)  \\
17  & hard   & 40.4 & 50.4 (61) & 100.0 (95) & 100.0 (95) \\
18  & hard   & 22.7 & 100.0 (99)& 100.0 (99) & 100.0 (99) \\
19  & hard   & 11.0 & 35.5 (82) & 100.0 (100)& 100.0 (100)\\
20  & hard   & 20.5 & 55.8 (89) & 94.5 (99)  & 94.6 (99)  \\
21  & medium & 20.3 & 100.0 (100)& 100.0 (100)& 100.0 (100)\\
22  & hard   & 22.1 & 49.6 (81) & 92.1 (99)  & 90.8 (99)  \\
23  & hard   & 13.6 & 100.0 (100)& 100.0 (100)& 100.0 (100)\\
24  & medium & 35.0 & 44.4 (62) & 100.0 (98) & 100.0 (98) \\
25  & hard   & 33.3 & 86.8 (94) & 95.3 (96)  & 100.0 (97) \\
26  & hard   & 22.2 &  8.3 (32) & 59.6 (89)  & 100.0 (99) \\
27  & hard   & 36.5 & 66.4 (80) & 74.7 (86)  & 99.5 (96)  \\
28  & medium & 30.3 & 54.1 (76) & 100.0 (98) & 100.0 (98) \\
29  & hard   & 20.3 & 66.8 (94) & 100.0 (100)& 98.0 (100) \\
30  & hard   & 35.0 & 61.2 (77) & 96.9 (96)  & 100.0 (97) \\
31  & hard   & 16.6 & 67.2 (96) & 98.4 (100) & 99.9 (100) \\
32  & hard   & 26.2 & 96.3 (98) & 98.5 (98)  & 100.0 (99) \\
33  & hard   & 15.2 & 68.8 (97) & 97.8 (100) & 98.9 (100) \\
34  & medium & 27.5 & 52.4 (80) & 99.5 (99)  & 99.3 (99)  \\
35  & hard   & 16.1 & 33.4 (74) & 89.3 (100) & 100.0 (100)\\
36  & hard   & 18.1 & 100.0 (100)& 100.0 (100)& 100.0 (100)\\
37  & hard   & 17.4 & 27.9 (65) & 100.0 (100)& 62.8 (95)  \\
38  & hard   & 20.3 & 91.8 (98) & 100.0 (99) & 100.0 (99) \\
39  & hard   & 36.0 & 61.8 (80) & 100.0 (98) & 98.4 (98)  \\
40  & hard   & 20.1 & 99.6 (99) & 100.0 (99) & 100.0 (99) \\
41  & hard   & 20.0 & 53.8 (86) & 97.0 (99)  & 96.9 (99)  \\
42  & medium & 29.5 & 58.9 (80) & 98.2 (98)  & 100.0 (98) \\
43  & hard   & 34.4 & 49.0 (67) & 98.2 (97)  & 68.8 (85)  \\
44  & hard   & 24.4 & 38.3 (65) & 100.0 (98) & 98.4 (98)  \\
45  & hard   & 12.3 & 14.8 (54) & 74.7 (99)  & 54.7 (95)  \\
46  & hard   & 21.7 & 28.6 (59) & 81.5 (97)  & 98.6 (99)  \\
47  & hard   & 28.0 & 35.6 (60) & 95.7 (99)  & 98.8 (99)  \\
48  & hard   & 12.9 & 55.6 (95) & 83.8 (100) & 94.0 (100) \\
49  & hard   & 14.3 & 57.6 (92) & 76.5 (98)  & 86.7 (99)  \\
50  & hard   & 20.0 & 17.3 (47) & 58.8 (89)  & 97.0 (99)  \\
\end{longtable}
\endgroup

\clearpage
\section*{Acknowledgment}

The authors thank Ilya Tillis of Codility for valuable guidance on experimental design.


\ifCLASSOPTIONcaptionsoff
  \newpage
\fi

\clearpage

\bibliographystyle{IEEEtran}
\input{output.bbl}

\end{document}